\documentclass{article}
\usepackage{graphicx} 
\usepackage{enumitem}
\usepackage{hyperref}
\usepackage{xcolor}
\usepackage{multirow,amsmath,amssymb}
\usepackage[T1]{fontenc}
\usepackage{bm}
\usepackage{setspace}
\usepackage{anysize}
\usepackage{cite}
\usepackage{url}
\marginsize{2cm}{2cm}{2cm}{1.5cm} 

\begin{document}
{ 
\allowdisplaybreaks
\ifdefined\rect \renewcommand{\rect}[1]{{\textstyle \frac{1}{#1}}}
\else             \newcommand{\rect}[1]{{\textstyle \frac{1}{#1}}}
\fi

\ifdefined\fract \renewcommand{\fract}[2]{{\textstyle \frac{#1}{#2}}}
\else             \newcommand{\fract}[2]{{\textstyle \frac{#1}{#2}}}
\fi

\ifdefined\soint \renewcommand{\soint}[4]{\hspace{#1}\oint_{#3}^{#4}\hspace{#2}}
\else              \newcommand{\soint}[4]{\hspace{#1}\oint_{#3}^{#4}\hspace{#2}}
\fi

\ifdefined\sointl \renewcommand{\sointl}[4]{\hspace{#1}\oint\limits_{#3}^{#4}\hspace{#2}}
\else               \newcommand{\sointl}[4]{\hspace{#1}\oint\limits_{#3}^{#4}\hspace{#2}}
\fi

\ifdefined\shointl \renewcommand{\shointl}[3]{\hspace{#1}\oint\limits^{#3}\hspace{#2}}
\else                \newcommand{\shointl}[3]{\hspace{#1}\oint\limits^{#3}\hspace{#2}}
\fi

\ifdefined\fracd
\renewcommand{\fracd}[2]{\frac{\displaystyle{#1}}{\displaystyle{#2}}}
\else
\newcommand{\fracd}[2]{\frac{\displaystyle{#1}}{\displaystyle{#2}}}
\fi

\ifdefined\bs \renewcommand{\bs}[1]{\boldsymbol{#1}}
\else            \newcommand{\bs}[1]{\boldsymbol{#1}}
\fi

\ifdefined\Eq \renewcommand{\Eq}[1]{Eq.~(\ref{#1})}
\else            \newcommand{\Eq}[1]{Eq.~(\ref{#1})}
\fi

\ifdefined\Eqs \renewcommand{\Eqs}[2]{Eqs.~(\ref{#1}) and (\ref{#2})}
\else            \newcommand{\Eqs}[2]{Eqs.~(\ref{#1}) and (\ref{#2})}
\fi

\ifdefined\sint   \renewcommand{\sint}[4]{\hspace{#1}\int_{#3}^{#4}\hspace{#2}}
\else               \newcommand{\sint}[4]{\hspace{#1}\int_{#3}^{#4}\hspace{#2}}
\fi

\ifdefined\rec   \renewcommand{\rec}[1]{\frac{1}{#1}}
\else              \newcommand{\rec}[1]{\frac{1}{#1}}
\fi

\ifdefined\z   \renewcommand{\z}[1]{\left({#1}\right)}
\else            \newcommand{\z}[1]{\left({#1}\right)}
\fi

\ifdefined\m   \renewcommand{\m}[1]{\mathrm{#1}}
\else            \newcommand{\m}[1]{\mathrm{#1}}
\fi

\ifdefined\v   \renewcommand{\v}[1]{\mathbf{#1}}
\else            \newcommand{\v}[1]{\mathbf{#1}}
\fi

\ifdefined\c   \renewcommand{\c}[1]{\mathcal{#1}}
\else            \newcommand{\c}[1]{\mathcal{#1}}
\fi

\setlist[itemize]{leftmargin=*}
\setlist[enumerate]{leftmargin=*}

\title{A novel method for calculating Bose-Einstein correlation functions 
with Coulomb final-state interaction
\footnote{
Dedicated to T.~Cs\"org\H o on the occasion of his 60th birthday.
}}
\author{M\'arton Nagy$^1$, Aletta Purzsa$^1$, M\'at\'e Csan\'ad$^1$, D\'aniel Kincses$^1$\footnote{kincses@ttk.elte.hu}\\$^1$E\"otv\"os Lor\'and University\\H-1117 Budapest, P\'azm\'any P\'eter s\'et\'any 1/A}

\date{July 2023}

\maketitle
\begin{abstract}
Measurements of Bose-Einstein correlations played a crucial role in the discovery and the subsequent detailed exploration of the Quark-Gluon-Plasma (QGP) created in high-energy collisions of heavy nuclei. Such measurements gave rise to femtoscopy, a flourishing sub-field of high-energy physics, and there are important new directions to explore and discoveries to be made in the near future. One of these important current topics is the precise investigation of the shape of the correlation functions utilizing L\'evy-stable sources. In this paper, we present a novel method of calculating the shape of the two-particle correlation functions, including the Coulomb final-state interaction. This method relies on an exact calculation of a large part of the necessary integrals of the Coulomb wave function and can be utilized to calculate the correlation function for any source function with an easily accessible Fourier transform. In this way, it is eminently applicable to L\'evy-stable source functions. In this particular case, we find that the new method is more robust and allows the investigation of a wider parameter range than the previously utilized techniques. We present an easily applicable software package that is ready to use in experimental studies.
\end{abstract}

\section{Introduction}\label{s:intro}

High energy heavy-ion physics aims at recreating the conditions characteristic to the first few microseconds of our Universe and studying the matter that is present under such conditions: the strongly interacting Quark Gluon Plasma (QGP)~\cite{STAR:2005gfr,Adcox:2004mh,PHOBOS:2004zne,BRAHMS:2004adc}. Experiments observe the particles produced in heavy ion collisions in order to draw conclusions about the properties of this matter. An important class of observables is that of momentum correlations of identical particles. In the case of bosonic particles (e.g., pions or kaons), these are called Bose-Einstein-correlations because they arise as a result of the quantum statistical properties of the particles. They are also called HBT correlations after R. Hanbury Brown and R. Q. Twiss~\cite{HanburyBrown:1956bqd}, who first discovered a related effect in astronomy. These correlations are connected to the space-time geometry of the particle emitting source~\cite{Lisa:2005dd}; thus, they provide a vital tool to study the matter produced in these collisions. These types of measurements have been extensively studied for more than 60 years now~\cite{Goldhaber:1960sf}, especially since the era of the Relativistic Heavy Ion Collider (RHIC) started around the early 2000s~\cite{PHENIX:2004yan,STAR:2004qya}. 

The usual way to experimentally study the source function is to assume a source shape (e.g., a Gaussian distribution) characterized by a handful of parameters, calculate the correlation function arising from such a source, then test this assumption on the measured momentum correlation function. If the calculated correlation shape fits the data well (in terms of statistical acceptability), one can extract the source parameters and investigate their dependence on average momentum, centrality, collision energy, etc. One of the most essential steps in such an analysis is thus the calculation of the correlation function for a given source function.

As detailed below, if one neglects the final state interactions of the produced particles, the correlation function is given by the Fourier transform of the source function, a relatively simple calculational step in most cases. However, in the case of charged particles (such as charged pions, the most frequent target of measurements), the correlation function is given by a more complicated integral transform of the source function. In the process of fitting the source parameters to measurements, a frequently used method has been to calculate the Coulomb integral only for a pre-defined set of the parameter values and derive a ``Coulomb correction factor'' from it. In this way, however, there is an inherent distortion in the fitting procedure since the Coulomb effect is calculated at some arbitrary parameter values instead of the supposedly final, fitted ones. As experimental measurements and phenomenological descriptions become more and more refined, it becomes absolutely necessary that in the fitting procedure, the Coulomb interaction is considered in a self-consistent way. This necessitates the calculation of the mentioned Coulomb transform of the assumed source function for arbitrary source parameters during the fitting procedure.

The simplest assumption for the source function is that of a Gaussian distribution, and in the past decades, many aspects of Gaussian correlation measurements were explored~\cite{Lisa:2000ip,Csanad:2008af,Pratt:2008qv}. However, as the experimental resolution and the accumulated data increased throughout the years, an expectation arose to better describe the correlation function's shape~\cite{PHENIX:2006nml}. One possibility is presented by a spherical harmonics expansion~\cite{Kisiel:2009iw}; another is to use a functional form for the source that goes beyond the Gaussian approximation. To that end, L\'evy-stable distributions were utilized~\cite{PHENIX:2017ino,NA61SHINE:2023qzr,CMS:2023xyd}. In this way, a statistically acceptable successful description of the data was achieved. The utilization of more and more general source function assumptions, such as the mentioned L\'evy distributions, is thus another pressing motivation for developing precise calculational tools to treat the Coulomb interaction.

In this paper, we present a new method for the calculation of the Bose-Einstein correlation function with the Coulomb interaction taken into account that is applicable to a wide variety of source functions. The new technique relies on the Fourier transform of the source function and is superior to already known methods in several ways: computationally, it is much cheaper and more stable than existing methods, and conceptually it highlights the actual effect of the Coulomb interaction on the correlation function in a straightforward way. In this paper, we present the method only for spherically symmetric source functions; however, the generalization to the non-spherical case is also being worked out. Another plausible future generalization is taking more complicated final state interaction schemes, such as strong interaction, into account, along the lines of Ref.~\cite{Kincses:2019rug}.

The structure of this paper is as follows. In Section~\ref{s:coulomb}, we detail the basic definitions and known methodology pertaining to Bose-Einstein correlations and the effect of the Coulomb interaction, paying particular attention to the role of L\'evy distributions as source functions. In Section~\ref{s:newmethod}, we present the new method. It has some delicate mathematical intricacies that are interesting on their own (so apart from the practical benefits of our new method, its derivation might interest the purely mathematically inclined reader as well). Some mathematical details are highlighted in Section~\ref{s:newmethod}, and some are left to the Appendix. In Section~\ref{s:results}, we verify our new method for the case of L\'evy-stable source functions. We find that the technique is significantly helpful and augments experimental measurements by making the correlation function fitting procedure easier. We also present a software package ready to use for such measurements~\cite{CoulCorrLevyIntegral}. Section~\ref{s:summary} summarizes our findings, pointing to possible future applications and generalizations.

\section{Bose-Einstein correlation functions and the Coulomb interaction}\label{s:coulomb}

The definition of the observable two-particle correlation function in a heavy-ion collision experiment is as follows: 
\begin{align}
\label{e:c2def}
C_2(\bs p_1,\bs p_2)=\frac{N_2(\bs p_1,\bs p_2)}{N_1(\bs p_1)N_1(\bs p_2)},
\end{align}
where $\bs p_1$ and $\bs p_2$ are the particle momentum (three-)vectors, $N_1(\bs p)\,{\equiv}\,E\frac{\m dn}{\m d^3\bs p}$ is the single-particle invariant momentum distribution (meaning the number of particles produced within a given invariant volume element in momentum space, and $E$ is the particle energy corresponding to momentum $\bs p$), and $N_2(\bs p_1,\bs p_2)\,{\equiv}\,E_1E_2\frac{\m d^2n}{\m d^3\bs p_1\,\m d^3\bs p_2}$ is the two-particle invariant momentum distribution, i.e., the number of particle \textit{pairs} produced with momenta $\bs p_1$ and $\bs p_2$ (and so with energies $E_1$ and $E_2$).

Before proceeding, let us introduce some appropriate notations and variables. Three-vectors are denoted by boldface letters, four-vectors by standard typeset letters; e.g., $\bs p_1$ is the three-momentum, $p_1\,{\equiv}\,(E_1,\bs p_1)$ is the four-momentum of particle 1. Lorentz products of four-vectors are denoted by a dot; for the metric tensor we use the $g^{\mu\nu}\,{=}\,\m{diag}(1,{-}1,{-}1,{-}1)$ convention. The scalar product (Descartes inner product) of three-vectors is denoted by writing the vectors next to each other. Below we deal with functions of two sets of space-time coordinates, $x_1\,{\equiv}\,(t_1,\bs r_1)$ and $x_2\,{\equiv}\,(t_2,\bs r_2)$; instead of these coordinates we can use the center-of-mass coordinate $X\,{\equiv}\,(T,\bs R)$ and the relative coordinates $x\,{\equiv}\,(t,\bs r)$ defined as
\begin{align}
\begin{array}{l}
X\,{:=}\,\rec2(x_1\,{+}\,x_2),\vspace{2pt}\\
x\;{:=}\,x_1\,{-}\,x_2,
\end{array}\qquad\Leftrightarrow\qquad
\begin{array}{l}
T\,{:=}\,\rec2(t_1\,{+}\,t_2),\vspace{2pt}\\
t\;{:=}\,t_1\,{-}\,t_2,
\end{array}\quad
\begin{array}{l}
\bs R\,{:=}\,\rec2(\bs r_1\,{+}\,\bs r_2),\vspace{2pt}\\
\bs r\;{:=}\,\bs r_1\,{-}\,\bs r_2.
\end{array}
\end{align}
Similarly, besides the momenta of the particles, $p_1$ and $p_2$, we use the average momentum four-vector $K$, as well as the relative momentum four-vector $Q$, and use the three-vector slices of these, $\bs K$ and $\bs Q$:
\begin{align}
    K\,{:=}\,\rec2(    p_1{+}\,    p_2),\qquad     Q :=     p_1{-}\,    p_2,\qquad
\bs K\,{:=}\,\rec2(\bs p_1{+}\,\bs p_2),\qquad \bs Q := \bs p_1{-}\,\bs p_2.
\end{align}
Since we are dealing with identical particles, their masses are equal, $m_1\,{=}\,m_2\,{\equiv}\,m$. With respect to the momentum variables, we usually suppress the $\hbar$ reduced Planck constant, meaning that momentum vectors are also understood as wave number vectors if the dimensions of the quantities imply so; for example, we write a plane wave-like wave function of the $\bs r$ variable like $e^{i\bs Q\bs r}$ but measure $\bs Q$ in MeV/$c$ like a momentum variable.

The $C_2(\bs p_1,\bs p_2)$ correlation function introduced in \Eq{e:c2def} above equals unity if the particle production is uncorrelated. In high-energy nuclear collisions, there are a number of processes that lead to $C_2\,{\neq}\,1$, i.e., correlated particle production (collective flow, jets, resonance decays, etc.). For identical bosons (such as charged pions), the main source of correlation at low momentum difference is what is called Bose-Einstein correlation, a quantum mechanical correlation stemming from the indistinguishability of the particles. For the simplest theoretical treatment of this, one can introduce the source function $S(x,\bs p)\,{\equiv}\,S(t,\bs r,\bs p)$ as the probability of a particle produced at space-time point $x\,{\equiv}\,(t,\bs r)$ with momentum $\bs p$. With this, the invariant momentum distribution is readily expressed as
\begin{align}
\label{e:N1p}
N_1(\bs p) = \sint{-2pt}{-3pt}{}{}\m d^4x\,S(x,\bs p),
\end{align}
where $\m d^4x$ stands for $\m dt\,\m d^3\bs r$, i.e., the integration is taken over the whole space-time; however, usually $S$ has a finite support or at least a rapid decrease in space and in time. Note that for the space-time coordinates, we have used $t$ and $\bs r$ as independent variables. On the other hand, for the momentum variable, we only write out $\bs p$ explicitly, with $E$ omitted, since for a given type of particle with mass $m$, $E$ is already determined by $\bs p$ as $E\,{=}\,\sqrt{m^2{+}\bs p^2}$.

Alternatively, one can argue that the $\Psi^{(1)}_{\bs p}(x)$ single-particle wave function describing a particle with momentum $\bs p$ is a plane wave, and thus 
\begin{align}
N_1(\bs p) = \sint{-2pt}{-3pt}{}{}\m d^4x\,S(x,\bs p)\big|\Psi^{(1)}_{\bs p}(x)\big|^2 = \sint{-2pt}{-3pt}{}{}\m d^4x\,S(x,\bs p),
\qquad\textnormal{since}\qquad \Psi^{(1)}_{\bs p}(x)\,{=}\,e^{-ip\cdot x}\quad\Rightarrow\quad \big|\Psi^{(1)}_{\bs p}(x)\big|^2\,{=}\,1.
\end{align}
The essence of the description of the Bose-Einstein correlation lies in the statement that the analog expression of $N_2(\bs p_1,\bs p_2)$ that takes into account particle production at two distinct space-time points $x_1$ and $x_2$ must be written up utilizing the two-particle wave-function $\Psi^{(2)}_{\bs p_1,\bs p_2}(x_1,x_2)$:~\cite{Yano:1978gk}
\begin{align}
\label{e:yanokoonin}
N_2(\bs p_1,\bs p_2) = \sint{-2pt}{-3pt}{}{}\m d^4x_1\,\m d^4x_2\,|\Psi^{(2)}_{\bs p_1,\bs p_2}(x_1,x_2)|^2\cdot S(x_1,\bs p_1) S(x_2,\bs p_2),
\end{align}
and that for bosons, $\Psi^{(2)}_{\bs p_1,\bs p_2}(x_1,x_2)$ is symmetric in the $x_1$ and $x_2$ variables. If the final state interactions of the particles are neglected, the wave function is a symmetrized plane wave, denoted by $\Psi^{(2,0)}_{\bs p_1,\bs p_2}$:
\begin{align}
\Psi^{(2,0)}_{\bs p_1,\bs p_2}(x_1,x_2) = \rec{\sqrt2}\Big[e^{-ip_1\cdot x_1}e^{-ip_2\cdot x_2}\,{+}\,e^{-ip_1\cdot x_2}e^{-ip_2\cdot x_1}\Big].
\end{align}
For the modulus square of this wave function, one obtains
\begin{align}
\label{e:onepluscosPsi}
\big|\Psi^{(2,0)}_{\bs p_1,\bs p_2}(x_1,x_2)\big|^2 = 1+\m{cos}\big(Q\,{\cdot}\,(x_1{-}x_2)\big),
\end{align}
and from this, after some simplifications, one gets the following expression for the correlation function (where the $0$ superscript denotes that final state interactions are neglected): 
\begin{align}
\label{e:Cp1p2:free}
C^{(0)}_2(p_1,p_2)= 1+ \m{Re}\frac{\widetilde S(Q,\bs p_1)\widetilde S^*(Q,\bs p_2)}{\widetilde S(0,\bs p_1)\widetilde S^*(0,\bs p_2)} ,
\end{align}
with $\widetilde S(Q,\bs p)$ being the Fourier transform of the source function:
\begin{align}
\label{e:tildeSdef}
\widetilde S(Q,\bs p) = \sint{-2pt}{-3pt}{}{}\m d^4x\,S(x,\bs p)e^{-iQ\cdot x}.
\end{align}
Typically the correlation function is written up as a function of the average and relative momentum variables $K$ and $Q$, respectively, instead of $p_1$, $p_2$. The reason for this is that it turns out that the dependence on $Q$ is much stronger than on $K$, so one almost always makes the so-called smoothness approximation~\cite{Pratt:1997pw} that in \Eq{e:Cp1p2:free}, $p_1\approx p_2\approx K$ in the second argument of $\widetilde S$. One then has
\begin{align}
\label{e:C2expr}
C^{(0)}_2(\bs Q,\bs K) = 1+\frac{|\widetilde S(Q,\bs K)|^2}{|\widetilde S(0,\bs K)|^2},
\end{align}
and one can conceptually separate the $Q$ and $K$ dependence by assuming that the source function $S(x,\bs K)$ has a given functional form (e.g., a Gaussian) as a function of $x$, and the parameters of this functional form may depend on $\bs K$. By measuring the correlation function as a function of $\bs Q$ at different $\bs K$ values, one can reconstruct the $\bs K$-dependence of the source parameters.

We can express the $\Psi^{(2,0)}$ two-particle wave function in terms of the center-of-mass system variables and the relative variables as
\begin{align}
\Psi^{(2,0)}_{\bs Q,\bs K}(x,X) = e^{-2iK\cdot X}\cdot\psi^{(0)}_{\bs Q,\bs K}(x),
\end{align}
where
\begin{align}
\label{e:relativepsi0}
\psi^{(0)}_{\bs Q,\bs K}(x) = \rec{\sqrt2}\big[e^{-i\rec2Q\cdot x}+e^{i\rec2Q\cdot x}\big].
\end{align}
In this form, we see that the relative motion and the center-of-mass motion disentangle. In the case of non-vanishing final state interactions, the center-of-mass motion still retains its plane-wave $e^{-2iQX}$ form. However, the form of the wave function pertaining to the relative motion changes. In this case, the two-particle wave function (now with the 0 superscript omitted, owing to the presence of final state interaction) is
\begin{align}
\Psi^{(2)}_{\bs Q,\bs K}(x,X) = e^{-2iK\cdot X}\cdot\psi_{\bs Q,\bs K}(x)\qquad\Rightarrow\qquad \big|\Psi^{(2)}_{\bs Q,\bs K}(x,X)\big|^2 = \big|\psi_{\bs Q,\bs K}(x)\big|^2.
\end{align}
Here the $\psi_{\bs Q,\bs K}(x)$ part of the wave function describes the relative motion. It is symmetric to the $x\,{\leftrightarrow}\,{-}x$ exchange (corresponding to the $x_1\,{\leftrightarrow}\,x_2$ change), and its concrete form depends on the interaction. We see that only this function determines the object of interest, the modulus of the two-particle wave function: the plane wave corresponding to the free motion of the center of mass cancels.

With this notation, we may rewrite the expression of the two-particle momentum distribution and the (utilizing the smoothness approximation, $\bs p_1\approx\bs p_2\approx\bs K$, just as above) as 
\begin{align}
\label{e:yanokoonin2}
&N_2(\bs Q,\bs K) = \sint{-2pt}{-3pt}{}{}\m d^4x\,\m d^4X\,S\big(X{+}\rect2x_1,\bs K\big)S\big(X{-}\rect2x,\bs K\big)|\psi_{\bs Q,\bs K}(x)|^2,\\
&N_1(\bs K)N_1(\bs K) = \sint{-2pt}{-3pt}{}{}\m d^4x\,\m d^4X\,S\big(X{+}\rect2x_1,\bs K\big)S\big(X{-}\rect2x,\bs K\big).
\end{align}
So with a new definition, we have
\begin{align}
C_2(\bs Q,\bs K) = \fracd{\sint{-2pt}{-3pt}{}{}\m d^4x\,D(x,\bs K)|\psi_{\bs Q,\bs K}(x)|^2}{\sint{-2pt}{-3pt}{}{}\m d^4x\,D(x,\bs K)},
\label{e:C2D}
\end{align}
where we introduced the $D(x,\bs K)$ relative coordinate distribution or spatial correlation function as  
\begin{align}
\label{e:DrKdef}
D(r,\bs K) \equiv \sint{-2pt}{-3pt}{}{}\m d^4X\,S\big(X{+}\rect2x,\bs K\big)S\big(X{-}\rect2x,\bs K\big).
\end{align}
Note that while $S(x,\bs p)$ is not necessarily an even function of $x$, this $D(x,\bs K)$ function is even in $x$. Taking this into account, and that from \Eq{e:relativepsi0}, we have $|\psi^{(0)}_{\bs Q,\bs K}(x)|^2\,{=}\,1\,{+}\,\cos(Q\cdot x)$, we can write the correlation function in the case of no final state interaction as
\begin{align}
\label{e:C2D0}
C_2^{(0)}(\bs Q,\bs K) = 1 + \frac{\widetilde D(Q,\bs K)}{\widetilde D(0,\bs K)},\qquad\textnormal{where}\quad
\widetilde D(Q,\bs K) = \sint{-2pt}{-3pt}{}{}\m d^4x\,D(x,\bs K)e^{-iQ\cdot x}.
\end{align}
We see that owing to the presence of the denominator in \Eqs{e:C2D}{e:C2D0}, the overall normalization of $D$ (the integral over space-time, at any given $\bs K$) cancels from the measurable $C_2$ correlation function. To simplify our formulas, in the following, we thus assume that $D$ is normalized to unity at any given $\bs K$.

For actual calculations, it would be desirable to have the time variable (in an appropriate reference frame) canceled from the production described by $S(x,\bs p)$. The freeze-out duration $\Delta\tau$ is in principle not zero (indeed, a prime goal of HBT studies is to measure $\Delta\tau$, and thus the order of phase transition). However, as multiple calculations show (see, e.g., Ref.~\cite{Rischke:1996em} for an early exposition), the effect of $\Delta\tau$ being non-zero can essentially be factored into the geometric (space-like) radii of the source function (with the time component of the relative momentum, $q^0$ being expressed with the spatial components along the way). So in the following, we write our formulas as if $D(x,\bs K)$ was non-zero only if $t{=}0$, highlighting this by omitting the time-dependence in the notation of our quantities, and remember that this is understood in the sense that the effect of finite time particle emission was treated by incorporating its effects into the geometrical radii of the source function. We thus have
\begin{align}
C_2(\bs Q,\bs K) = \sint{-2pt}{-3pt}{}{}\m d^3\bs r\,D(\bs r,\bs K)|\psi_{\bs Q,\bs K}(\bs r)|^2,
\end{align}
specifically, for the case when final state interactions are neglected,
\begin{align}
\label{e:C0D1}
C^{(0)}_2(\bs Q,\bs K) = 1 + \sint{-2pt}{-3pt}{}{}\m d^3\bs r\,D(\bs r,\bs K)e^{i\bs Q\bs r}.
\end{align}
We also appropriately change our notation at this point: since we no longer need four-vectors, in the case of a three-vector (denoted by bold letter), the same letter in standard typeset indicates the magnitude of the vector, e.g., $Q\equiv |\bs Q|$, $r\equiv|\bs r|$.

In the case of the Coulomb interaction, the $\psi(x)$ wave function (a solution to the Schrödinger equation satisfying appropriate boundary conditions) has a known form in the non-relativistic case. This implies that one has to use the center-of-mass frame of the particle pair (PCMS frame), where their motion can be approximated to be non-relativistic. In the following thus, we take $\bs Q$ to be understood in this PCMS frame and use the $\bs k\,{=}\,\bs Q/2$ notation, and (as customary for such treatment of the Coulomb interaction) assume that the simultaneity condition expounded above holds in PCMS. So, in this case, we have
\begin{align}
C_2(\bs Q) = \sint{-2pt}{-3pt}{}{}\m d^3\bs r\,D(\bs r)|\psi_{\bs k}(\bs r)|^2,
\label{e:c2final}
\end{align}
where we dropped the $\bs K$ arguments for simplicity, but remember that the source function depends on $\bs K$ through the $\bs K$-dependence of its assumed parameters, and use the appropriately symmetrized $\psi_{\bs k}(\bs r)$ Coulomb interacting \textit{out} state in the following form, well known in quantum mechanical scattering theory (see, e.g., Ref.~\cite{Landau:1991wop} for details):
\begin{align}
\label{e:psikr}
\psi_{\bs k}(\bs r) = \frac{\mathcal N^*}{\sqrt2}e^{-ikr}\Big[M\big(1{-}i\eta,1,i(kr{+}\bs k\bs r)+M\big(1{-}i\eta,1,i(kr{-}\bs k\bs r)\big)\Big].
\end{align}
The $\eta$ quantity appearing here is the so-called Sommerfeld parameter:
\begin{align}
\eta:=\frac{mc^2\alpha}{2\hbar kc},
\end{align}
with $\alpha\equiv\frac{q_e^2}{4\pi\varepsilon_0}\rec{\hbar c}\approx\rec{137}$ being the fine structure constant. The quantity $\c N$ is the normalization factor of the wave function, of whose square modulus is the so-called \textit{Gamow factor}:
\begin{align}
\c N = e^{-\pi\eta/2}\Gamma(1{+}i\eta)\qquad\Rightarrow\qquad |\c N|^2 = \frac{2\pi\eta}{e^{2\pi\eta}\,{-}\,1},
\end{align}
where $\Gamma(z)$ is the Gamma function, and its simple properties (such as $z\Gamma(z)\,{=}\,\Gamma(z{+}1)$ and $\Gamma(z)\Gamma(1{-}z)=\frac\pi{\sin(\pi z)}$) lead to the expression written up for $|\c N|^2$. Finally, $M(a,b,z)$ is the confluent hypergeometric function,
\begin{align}
\label{e:confhypdef}
M(a,b,z) = 1 + \frac ab\frac{z}{1!} + \frac{a(a{+}1)}{b(b{+}1)}\frac{z^2}{2!}+\ldots = \sum_{n=0}^\infty\frac{\Gamma(a{+}n)}{\Gamma(a)}\frac{\Gamma(b)}{\Gamma(b{+}n)}\frac{z^n}{n!}.
\end{align}

Besides the (modulus square of the) wave function, an assumption is needed about the functional form of the $D(\bs r)$ pair distribution in order to calculate the correlation function as in \Eq{e:c2final}. As mentioned in Section~\ref{s:intro}, recent measurements utilized L\'evy-stable distributions as an assumption for $D(\bs r)$. The appearance of Lévy-stable sources in high energy heavy-ion collisions was expounded in~\cite{Csorgo:2003uv}. Since then, several mechanisms were proposed as the possible cause for this particular source shape, such as jet fragmentation~\cite{Csorgo:2004sr}, critical phenomena~\cite{Csorgo:2005it}, directional averaging and non-sphericality~\cite{Cimerman:2019hva}, event averaging~\cite{Cimerman:2019hva}, resonance decays and hadronic rescattering~\cite{Csanad:2007fr,Kincses:2022eqq}. Out of these, each may result in Lévy-shaped sources at different collision energies and systems: jet fragmentation in e$^+$e$^{-}$ or pp collisions, critical phenomena at energies lower than the top RHIC or LHC energies, while hadronic rescattering may occur generally. In~\cite{Kincses:2022eqq}, it was shown that in realistic simulations of the hadronic medium, L\'evy sources appear on an event-by-event basis at various stages of the evolution.

The symmetric L\'evy distribution is a generalization of the Gaussian function as it has an additional parameter $\alpha$, called the L\'evy exponent. In the spherically symmetric case, it has the following form:
\begin{align}
\label{e:Levy3Ddef}
\c L(\alpha,\bs R,\bs r):=\rec{(2\pi)^3}\int\m d^3\bs q\, e^{i\bs q\bs r} e^{-\frac{1}{2}|\bs qR|^{\alpha}}.
\end{align}
By the known property of Fourier transforms, one sees that this function is normalized to unity:
\begin{align}
\sint{-2pt}{-3pt}{}{}\m d^3\bs r\,\c L(\alpha,R,\bs r) = 1.
\end{align}
In the $\alpha\,{=}\,2$ case we recover a Gaussian distribution with radius $R$; while for $\alpha\,{=}\,1$, a Cauchy distribution is obtained:
\begin{align}
\label{e:Levy:speccases}
\c L(\alpha{=}2,R,\bs r) = \rec{(2\pi R^2)^{3/2}}e^{-\frac{r^2}{2R^2}},\qquad\qquad
\c L(\alpha{=}1,R,\bs r) = \frac8{\pi^2R^3}\rec{\big(1+\frac{4r^2}{R^2}\big)^2}.
\end{align}
For $\alpha\,{\neq}\,2$ and $\alpha\,{\neq}\,1$, one cannot express $\c L(\alpha,R,r)$ with a combination of simple functions. However, its asymptotic expression is known: for every $\alpha\,{\neq}\,2$, it decreases like a power of $r$:
\begin{align}\label{e:r2Ldist}
\textnormal{for large $r$,}\quad r^2 \c L(\alpha,R,\bs r) \propto r^{-1-\alpha},
\end{align}
as exemplified in the above $\alpha{=}1$ case as well.

L\'evy distributions generalize Gaussian distributions also in the sense that they retain the stability property of Gaussian: the convolution of two L\'evy distributions (with the same $\alpha$ parameter) is again such a L\'evy function, as can be directly seen by inserting the above definition (a Fourier integral) and performing the integrals. Specifically,
\begin{align}
\sint{-2pt}{-3pt}{}{}\m d^3\bs\rho\,\c L\big(\alpha,R_1,\bs\rho{+}\rect2\bs r\big)\c L\big(\alpha,R_2,\bs\rho{-}\rect2\bs r\big) = \c L\big(\alpha,R,\bs r\big),\qquad\textnormal{with}\quad R^\alpha = R_1^\alpha +R_2^\alpha.
\label{e:levyconv1}
\end{align}
From another point of view, with a similar calculation, we obtain that if the $S(\bs r,\bs p)$ single-particle source function is a L\'evy-type function, then so is the $D(\bs r,\bs p)$ two-particle source function:
\begin{align}
S(\bs r,\bs p) = \c L\Big(\alpha(\bs p),R(\bs p),\bs r{-}\bs r_0(\bs p)\Big)
\qquad\Rightarrow\qquad
D(\bs r,\bs p) = \c L\Big(\alpha(\bs p),2^{1/\alpha(\bs p)}R(\bs p),\bs r\Big).
\label{e:levyconv2}
\end{align}
We could even allow $\bs p$-dependent parameters as well as an arbitrary $\bs r_0(\bs p)$  shift in $S(\bs r,\bs p)$; this does not change the shape of $D(\bs r,\bs p)$, the quantity that appears in the expression of the correlation function.

The stability property written up in \Eq{e:levyconv1} is the main reason why L\'evy distributions are expected to appear in many different circumstances (as limiting distributions, just as Gaussians). In the scenarios mentioned above, they also naturally arise as particle production source functions in heavy-ion collisions, the main object of interest in this paper. Indeed, L\'evy distributions have already been successfully applied to describe such sources~\cite{PHENIX:2017ino}, after earlier measurements indicated that a Gaussian description falls short of such data~\cite{PHENIX:2006nml}. However, as explained below, it turns out that the actual calculation of the $C_2$ correlation function for L\'evy-like $D(\bs r,\bs p)$ is cumbersome even numerically; there is thus a natural need for the development of such calculational methods. This becomes even more pressing if one abandons spherical symmetry or if one allows further generalizations of the source function beyond the L\'evy assumption.

\section{A new method for the treatment of the Coulomb effect}\label{s:newmethod}

As written up in Eq.~(\ref{e:c2final}) above, for calculating the correlation function with the Coulomb interaction taken into account, one has to integrate the modulus of the Coulomb interacting wave function $\psi_{\bs k}(\bs r)$ weighted with the $D(\bs r)$ two-particle source function. This is deemed to be feasible only numerically, owing to the complicated form of $\psi_{\bs k}(\bs r)$, Eq.~(\ref{e:psikr}). Nevertheless, as seen above in \Eq{e:C0D1}, in case of vanishing Coulomb force, $|\psi_{\bs k}(\bs r)|$ is a plane wave, and our desired integral reduces to a Fourier transformation of $D(\bs r)$.

There is an important, if not the only interesting, class of possible source function assumptions that have the property of being defined and best calculable as a Fourier transform of some simple function. A Gaussian source function obviously falls into this category; however, a preeminent motivating example is when the $D(\bs r)$ two-particle source function is a L\'evy distribution, whose Fourier transform, $e^{-|R\bs q|^\alpha}$ is an easily computable fast decreasing function. In the following, we thus assume that
\begin{align}
\label{e:Dwithf}
D(\bs r) = \rec{(2\pi)^3}\sint{-2pt}{-2pt}{}{}\m d^3\bs q\,f(\bs q)e^{i\bs q\bs r}.
\end{align}
We assume that $f(\bs q)$ is an integrable function (over the whole $\bs q$ space); this is a natural assumption if we want to interpret this Fourier transform as a regular (Lebesgue-)integral. We also assume $D(\bs r)$ to be an integrable function (over the whole $\bs r$ space; this is naturally fulfilled whenever $S(\bs r)$ is integrable, which is necessary to interpret $S(\bs r)$ as the function whose integral, according to \Eq{e:N1p}, gives the single particle invariant momentum distribution). These imply that both $f(\bs q)$ and $D(\bs r)$ are bounded and continuous, as well as that the (inverse) Fourier transform of $D(\bs q)$ is also a regular integral:
\begin{align}
\label{e:fwithD}
f(\bs q) = \sint{-2pt}{-2pt}{}{}\m d^3\bs r\,D(\bs r)e^{-i\bs q\bs r},
\end{align}
and thus the normalization condition for $D(\bs r)$ means a simple condition on $f$:
\begin{align}
\sint{-2pt}{-3pt}{}{}\m d^3\bs r\,D(\bs r) = 1 \qquad\Leftrightarrow\qquad
f(\bs q{=}0) = 1.
\end{align}
In particular, the interaction-free correlation function has $f$ as the main component; from \Eqs{e:C0D1}{e:fwithD},
\begin{align}
\label{e:C0fromf}
C^{(0)}_2(\bs Q) = 1 + f(\bs Q).
\end{align}

In the Coulomb interacting case, our goal is to calculate the correlation function using \Eq{e:c2final}; put together with \Eq{e:Dwithf}, we have
\begin{align}
C_2(\bs Q) = \rec{(2\pi)^3}\sint{-2pt}{-3pt}{}{}\m d^3\bs r\,|\psi_{\bs k}(\bs r)|^2\sint{-2pt}{-3pt}{}{}\m d^3\bs q\,f(\bs q)e^{i\bs q\bs r},
\label{e:CDf0}
\end{align}
where $\psi_{\bs k}(\bs r)$ is the Coulomb wave function written up in \Eq{e:psikr}. The straightforward way to proceed used, e.g., in Refs.~\cite{Csanad:2019cns,Kincses:2019rug} in the case of L\'evy distributions, is then to calculate $D(\bs r)$ from $f(\bs q)$ as a Fourier integral, then perform the integral over $\bs r$, in which $|\psi_{\bs k}(\bs r)|^2$ enters. Numerically, this is a daunting task in some cases. E.g., for the L\'evy distribution, the result of the Fourier transform (the function $D(\bs r)$ itself) is only slowly decreasing (with a power-law-like behavior for large $r$), while $|\psi_{\bs k}(\bs r)|^2$ is an oscillating function; asymptotically a plane wave (up to logarithmic corrections). There is also a conceptual awkwardness in this methodology: we take a Fourier transform of $f(\bs q)$ and then subject the resulting $D(\bs r)$ function to an ``almost inverse Fourier transform'' (i.e., the $\bs r$-integral with the ``almost plane wave'' $|\psi_{\bs k}(\bs r)|^2$ as a kernel) to finally arrive at $C_2(\bs Q)$. In the interaction-free case, $|\psi_{\bs k}(\bs r)|^2$ is really a plane wave, and the result, \Eq{e:C0fromf} indeed shows that these two Fourier transformations cancel each other. One would thus very much prefer a calculational scheme where this ``back and forth'' transformation is not needed in its full numerical complexity. (For example, calculating $C_2(\bs Q)$ at high $\bs Q$ values, where owing to the decrease of $f(\bs q)$ it is increasingly closer to unity, still requires significant computational power because at higher $k$, the oscillations of $\psi_{\bs k}(\bs r)$ become faster and faster.) 

The natural idea to resolve these problems is that we would like to perform the Fourier transform of the interacting $|\psi_{\bs k}(\bs r)|^2$ function and ``act'' with this resulting integral kernel (a function of $\bs k$ and $\bs q$) on the $f(\bs q)$ function. In other words, we would like to interchange the order of integrals in \Eq{e:CDf0} so that we could write
\begin{align}
C_2(\bs Q)\stackrel{??}\leftrightarrow \rec{(2\pi)^3}\sint{-2pt}{-3pt}{}{}\m d^3\bs q\,f(\bs q)\sint{-2pt}{-3pt}{}{}\m d^3\bs r\,e^{i\bs q\bs r}|\psi_{\bs k}(\bs r)|^2.
\label{e:Cfpsi:wrong}
\end{align}
Regrettably, however, this is not possible in such a simple form: since $\psi_{\bs k}(\bs r)$ is asymptotically a plane wave, its modulus is definitely not an integrable function whose Fourier transform could simply be calculated by an integral. In the following, we work around this problem by carefully treating integrability and the interchange of integrals and limits, a rarely found exercise in physics-motivated standard calculations. The resulting formulas, however, are worth such careful investigation. In doing so, we utilize some fundamental theorems about (Lebesgue) integrability.%
\footnote{
These theorems provide the framework for the practical applications of the theory of distributions. Indeed, the main idea behind our calculation is deeply connected to the theory of distributions, however, we do not explicitly make use of any strictly distribution theory results. 
}
We outline the main steps of our calculations below but leave some mathematical and calculational details to the Appendix.

It turns out that we can proceed by inserting an exponential ``regularization'', $e^{-\lambda r}$ into the integrand of \Eq{e:CDf0}, with a positive real $\lambda\,{>}\,0$ parameter, which at the end goes to 0. With finite $\lambda$ we can interchange the order of integrals and finally arrive at
\begin{align}
C_2(\bs Q) &= \sint{-2pt}{-3pt}{}{}\m d^3\bs r\,|\psi_{\bs k}(\bs r)|^2\sint{-2pt}{-3pt}{}{}\frac{\m d^3\bs q}{(2\pi)^3}\,f(\bs q)e^{i\bs q\bs r}
=\nonumber\\& = \sint{-2pt}{-3pt}{}{}\m d^3\bs r\,\lim_{\lambda\to0}e^{-\lambda r}|\psi_{\bs k}(\bs r)|^2\sint{-2pt}{-3pt}{}{}\frac{\m d^3\bs q}{(2\pi)^3}\,f(\bs q)e^{i\bs q\bs r}
\stackrel{1.}= \nonumber\\&\stackrel{1.}=\lim_{\lambda\to0}\sint{-2pt}{-3pt}{}{}\m d^3\bs r\,\sint{-2pt}{-3pt}{}{}\frac{\m d^3\bs q}{(2\pi)^3}\,e^{-\lambda r}|\psi_{\bs k}(\bs r)|^2\,f(\bs q)e^{i\bs q\bs r}
\stackrel{2.}= \nonumber\\&\stackrel{2.}=\lim_{\lambda\to0}\sint{-2pt}{-3pt}{}{}\frac{\m d^3\bs q}{(2\pi)^3}\,f(\bs q)\sint{-2pt}{-3pt}{}{}\m d^3\bs r\,e^{-\lambda r}|\psi_{\bs k}(\bs r)|^2\,e^{i\bs q\bs r}.
\label{e:CDf1}
\end{align}
(Remember that $\bs Q$ is the variable of the resulting observable $C_2$ correlation function, $\bs Q\,{=}\,2\bs k$, while $\bs q$ is a mere auxiliary integration variable in this calculation.)

In \Eq{e:CDf1}, every step is justified mathematically, in particular, the exchange of the limit and the integral in Step 1 (by virtue of \textit{Lebesgue's theorem}), and the exchange of the order of integrals in Step 2 (by virtue of \textit{Fubini's theorem}); see Appendix \ref{ss:app:lebesgue} for details. Nevertheless, the point is that in the resulting final formula, the $\lambda\,{\to}\,0$ limit \textit{cannot} be exchanged with the $\bs q$-integral. The way to proceed is to calculate the $\bs r$-integral for finite $\lambda$  values, and then ``simplify'' the $\lambda\,{\to}\,0$ limit to arrive at a final formula where there are no explicit limits that would have to be evaluated numerically (which would be very challenging if not impossible).

In this paper, we proceed with pair source functions that are spherically symmetric; this also implies spherical symmetry of $f(\bs q)$, which we highlight everywhere with an $s$ in the subscript. In the spherically symmetric case, we can perform the solid angle integral in the Fourier integral, arriving at the following expression of $D(\bs r)$:
\begin{align}
f(\bs q)\equiv f_s(q) \qquad\Rightarrow\qquad
D(\bs r) \equiv D(r) =
\rec{2\pi^2r}\sint{-2pt}{-10pt}0\infty\m dq\,q\sin(qr)f_s(q).
\end{align}
Following the exact same steps as in \Eq{e:CDf1}, with this expression for the spherically symmetric $D(r)$ we have
\begin{align}
C_2(Q) = \rec{2\pi^2}\lim_{\lambda\to0}\sint{-2pt}{-10pt}0\infty\m dq\,q^2f_s(q)\sint{-2pt}{-3pt}{}{}\m d^3\bs r\,e^{-\lambda r}\frac{\sin(qr)}{qr}|\psi_{\bs k}(\bs r)|^2.
\label{e:CDf1spher}
\end{align}
Note that while $\psi_{\bs k}(\bs r)$ depends on the direction of $\bs k\,{=}\,\bs Q/2$, the result of this integral indeed only depends on the magnitude $Q$, after performing the $\bs r$-integral.

As a next step, we substitute $\psi_{\bs k}(\bs r)$ from \Eq{e:psikr}. The modulus square results in four terms (from each pairings of the different $M(a,b,z)$ functions). With an appropriate $\bs r\,{\to}\,{-}\bs r$ change of variable in two of these, we are left with only two different terms:
\begin{align}
\label{e:CQspherD1D2}
&C_2(Q) = \frac{|\c N|^2}{2\pi^2}\lim_{\lambda\to0}\sint{-2pt}{-10pt}0\infty\m dq\,q^2f_s(q)\Big[\c D_{1\lambda s}(q)+\c D_{2\lambda s}(q)\Big],\qquad\textnormal{where}\\
&\qquad\c D_{1\lambda s}(q) = \sint{-3pt}{-2pt}{}{}\m d^3\bs r\frac{\sin(qr)}{qr}e^{-\lambda r}
M\big(1{+}i\eta,1,-i(kr{+}\bs k\bs r)\big)M\big(1{-}i\eta,1,i(kr{+}\bs k\bs r)\big), \label{e:D1sdef}\\
&\qquad\c D_{2\lambda s}(q) = \sint{-3pt}{-2pt}{}{}\m d^3\bs r\frac{\sin(qr)}{qr}e^{-\lambda r}
M\big(1{+}i\eta,1,-i(kr{-}\bs k\bs r)\big)M\big(1{-}i\eta,1,i(kr{+}\bs k\bs r)\big). \label{e:D2sdef}
\end{align}
The integrals of \Eqs{e:D1sdef}{e:D2sdef} are calculated in Appendix~\ref{ss:app:nord}; the utilized method is similar to that of Nordsieck~\cite{Nordsieck:1954zz}. In our case, the result is
\begin{align}
\label{e:D1lambdas}
&\c D_{1\lambda s}(q) = \frac{4\pi}q\m{Im}\bigg[
\rec{(\lambda{-}iq)^2}\Big(1{+}\fract{2k}{q{+}i\lambda}\Big)^{2i\eta}{}_2F_1\Big(i\eta,1{+}i\eta,1,\fract{4k^2}{(q{+}i\lambda)^2}\Big)\bigg],\\
&\c D_{2\lambda s}(q) = \frac{4\pi}q\m{Im}\bigg[
\frac{(\lambda{-}iq{-}2ik)^{i\eta}(\lambda{-}iq{+}2ik)^{-i\eta}}{(\lambda{-}iq)^2{+}4k^2}\bigg],
\label{e:D2lambdas}
\end{align}
where we make use of the (ordinary) hypergeometric function:
\begin{align}
{}_2F_1\big(a,b,c,z\big)=1+\frac{ab}c\frac{z}{1!} + \frac{a(a{+}1)b(b{+}1)}{c(c{+}1)}\frac{z^2}{2!}+\ldots = \sum_{n=0}^\infty\frac{\Gamma(a{+}n)\Gamma(b{+}n)}{\Gamma(a)\Gamma(b)}\frac{\Gamma(c)}{\Gamma(c{+}n)}\frac{z^n}{n!}.
\label{e:hypdefpowerseries}
\end{align}
This power series is valid only for $|z|\,{<}\,1$, however, the domain of ${}_2F_1(a,b,c,z)$ can be extended to any $z$ complex number arguments by analytic continuation; see e.g.~Ref.~\cite{NIST:DLMF}.

The next step is then to try to evaluate $C_2(Q)$ as in \Eq{e:CQspherD1D2} from these expressions of $\c D_{1\lambda s}$ and $\c D_{2\lambda s}$. For this, we need to take the $\lambda\,{\to}\,0$ limit in a careful way. It turns out that as $\lambda$ approaches 0, the functional forms of $\c D_{1\lambda s}$ and $\c D_{2\lambda s}$ become ill-behaved; the desired limit (of the result of the $q$-integral) cannot simply be expressed as some limiting $q$-function multiplied by $f_s(q)$ and then integrated. Instead, in the final expression, one has to take a specific linear \textit{functional} of $f_s(q)$.%
\footnote{
The situation is similar in the well-known much simpler case when one approximates the Dirac delta with ever narrower and higher peaks with area of unity. Consider the following identity:
\begin{align*}
\lim_{\lambda\to0}\sint{-0pt}{-12pt}{-\infty}\infty\m dx\,\rec\pi\frac\lambda{\lambda^2{+}(x{-}x_0)^2}f(x) = f(x_0).
\end{align*}
Based on this, one can say that the narrower and higher Lorentz curves ``approximate the $\delta(x{-}x_0)$ delta function''; the meaning of this statement is essentially the same as the displayed identity. From a practical point of view, the benefit of this formula is that instead of numerically performing some $\lambda\,{\to}\,0$ limit of the result of the left-hand side integral, the right-hand side provides a much simpler calculational statement (by requiring just to evaluate the $f$ function at a given $x_0$ point); however, this simpler statement is no longer a true integral transformation acting on $f$.
}
With leaving some intermediate steps to Appendix~\ref{ss:app:finallimit}, in the following Section, we begin by writing up the main result of our calculation for $C_2(Q)$.

\section{Results and discussion}\label{s:results}


\subsection{The new formula for $C_2(Q)$}

Starting from the expression of $C_2(Q)$ from \Eq{e:CQspherD1D2}, and substituting $\c D_{1\lambda s}$ and $\c D_{2\lambda s}$ from \Eqs{e:D1lambdas}{e:D2lambdas}, a careful utilization of \textit{Lebesgue's theorem} (detailed in Appendix~\ref{ss:app:finallimit}) allows one to perform the remaining $\lambda\,{\to}\,0$ limit in \Eq{e:CQspherD1D2}. We thus arrive at the main result of our new method for calculating the Coulomb interacting $C_2(Q)$, which can be summarized as follows:
\begin{align}
\label{e:CwithA1A2}
C_2(Q) = |\mathcal N|^2\bigg(1 + f_s(2k)+\frac\eta\pi\big[\c A_{1s}+\c A_{2s}\big]\bigg),
\end{align}
where the $\c A_{1s}$ and $\c A_{2s}$ terms are the following functionals of $f_s(q)$:
\begin{align}
\label{e:A1sexpr}
&\c A_{1s} = -\frac2\eta\sint{-2pt}{-10pt}0\infty\m dq\,\frac{f_s(q){-}f_s(0)}q\m{Im}\bigg[\bigg(1{+}\frac{2k}q\bigg)^{2i\eta}{}_2F_1\Big(i\eta,1{+}i\eta,1,\frac{4k^2}{q^2}{-}i0\Big)\bigg],\\
&\c A_{2s} = -\frac2\eta\sint{-2pt}{-10pt}0\infty\m dq\,\frac{f_s(q){-}f_s(2k)}{q{-}2k}\frac q{q{+}2k}\m{Im}\frac{(q{+}2k)^{i\eta}}{(q{-}2k{+}i0)^{i\eta}}.
\label{e:A2sexpr}
\end{align}
If $\eta\,{\to}\,0$ (either by letting $k\,{\to}\,\infty$, or by formally turning off the Coulomb force), $\c A_{1s}$ and $\c A_{2s}$ take on finite values (because both the $\eta$ denominator and the imaginary values of the denoted quantities go to 0). So for $\eta\,{\to}\,0$, in \Eq{e:CwithA1A2} the contributions of $\c A_{1s}$ and $\c A_{2s}$ vanish (where they are again multiplied by $\eta$); we wrote the $\eta$ factors in the way we did to highlight this feature.

Note also that in \Eq{e:CwithA1A2}, the effect of Coulomb interaction enters in a straightforward, ``traceable'' way. Omitting $|\c N|^2$ as well as the $\c A_{1s}$, $\c A_{2s}$ terms, and recalling that $\bs Q\,{=}\,2\bs k$, we simply have \Eq{e:C0fromf}, the result for the interaction-free case. Including the $|\c N|^2$ factor (but not $\c A_{1s}$ and $\c A_{2s}$) is the so-called \textit{Gamow correction}, the simplest approximate treatment of the Coulomb interaction: it treats the source as point-like for the Coulomb integration (but not for the calculation of the interaction-free correlation function). The $\c A_{1s}$ and $\c A_{2s}$ terms can thus be thought of as the effect of the source being not point-like; the parts of the correlation function that is neglected by the Gamow prescription.

In our normalization, $f_s(0)\,{=}\,1$; however, we retained $f_s(0)$ in \Eq{e:A1sexpr} to highlight that the fraction containing the $f_s$ function is a well-defined function even around $q\,{=}\,0$ (provided that $f_s$ is continuously differentiable). Similarly, the fraction containing $f_s$ in the expression of $\c A_{2s}$, \Eq{e:A2sexpr}, is also a well-defined function for continuously differentiable $f_s$, even around $q\,{=}\,2k$. In fact, we can loosen the requirement of $f_s$ being continuously differentiable. The more general assumption under which the above formulas are derived is explained (along with other details) in Appendix~\ref{ss:app:finallimit} around \Eqs{e:fsfracbounded}{e:powerpeak}. Here we just remark that this more general assumption is satisfied already if $f(\bs q)$ is everywhere continuously differentiable, and also in case of L\'evy distributions (for which $f(\bs q)$ is not differentiable at the origin if $\alpha\,{\le}\,1$).

Power functions of complex variables are understood in the strictly single-valued function sense: for $z,w\,{\in}\,\mathbb C$, $z^w\,{:=}\,\exp(w\,\m{Ln}\,z)$, where $\exp$ is an entire function, its inverse however, $\m{Ln}\,z\equiv \log(|z|)+i\m{arg} z$, has a branch cut along the $\mathbb R_0^-$ negative real line (owing to the jump of $\m{arg}\,z$, the phase of $z$, from $\pi$ to $-\pi$, when crossing this line in a counterclockwise direction). This branch cut along $\mathbb R_0^-$ is inherited by the power function $z^w$ as a function of $z$, whenever $w\,{\notin}\,\mathbb Z$. So when taking a power of a quantity that is on this branch cut, we have to specify which side of the cut we are on. This is the reason for the $+i0$ term in \Eq{e:A2sexpr}. Also when one defines the ${}_2F_1(a,b,c,z)$ hypergeometric function for $|z|\,{\ge}\,1$, that is, outside the convergence radius of the power series in \Eq{e:hypdefpowerseries}, one encounters a branching point at $z\,{=}\,1$, and the usual convention places the branch cut on the $[1,\infty]\,{\subset}\,\mathbb R^+$ real line. The $-i0$ term in \Eq{e:A1sexpr} specifies which side of this cut we are on.

\subsection{Utilizing the new formula for L\'evy-stable sources}\label{ss:results:calcforlevy}

Before delving into the details of the applied computational methods, in Figure~\ref{fig:compare}, we show the main results of our calculations, i.e., the $C_2(Q)$ correlation functions calculated for L\'evy-stable source distributions $\c L(\alpha,R,r)$, separately for pion-pion and kaon-kaon pairs. Different colors denote different L\'evy scale $R$ values, while the dashed and solid lines correspond to the L\'evy exponent $\alpha$ values of 0.6 and 2, respectively. Correlation functions with $\alpha$ between these two extremal values are illustrated with a filled area. In the interaction-free case, for any given $R$, correlation functions take the same value for any $\alpha$ at one specific $Q\,{>}\,0$ value: ${C_2^{(0)}(Q = \hbar c/R) = 1+1/e}$. In the case when Coulomb interaction is included, the curves with different $\alpha$ values still intersect each other at approximately the same point (for a given $R$), in the vicinity of $Q\approx\hbar c/R$. However, in this case, the $C_2(Q\approx\hbar c/R)$ values have an $R$ dependence. Moreover, the intersection of all the curves is only approximate (albeit a very good one); higher zooming around the apparent ``nodes'' would tell the curves apart.

\begin{figure}[h!]
    \centerline{
    \includegraphics[width=0.5\textwidth]{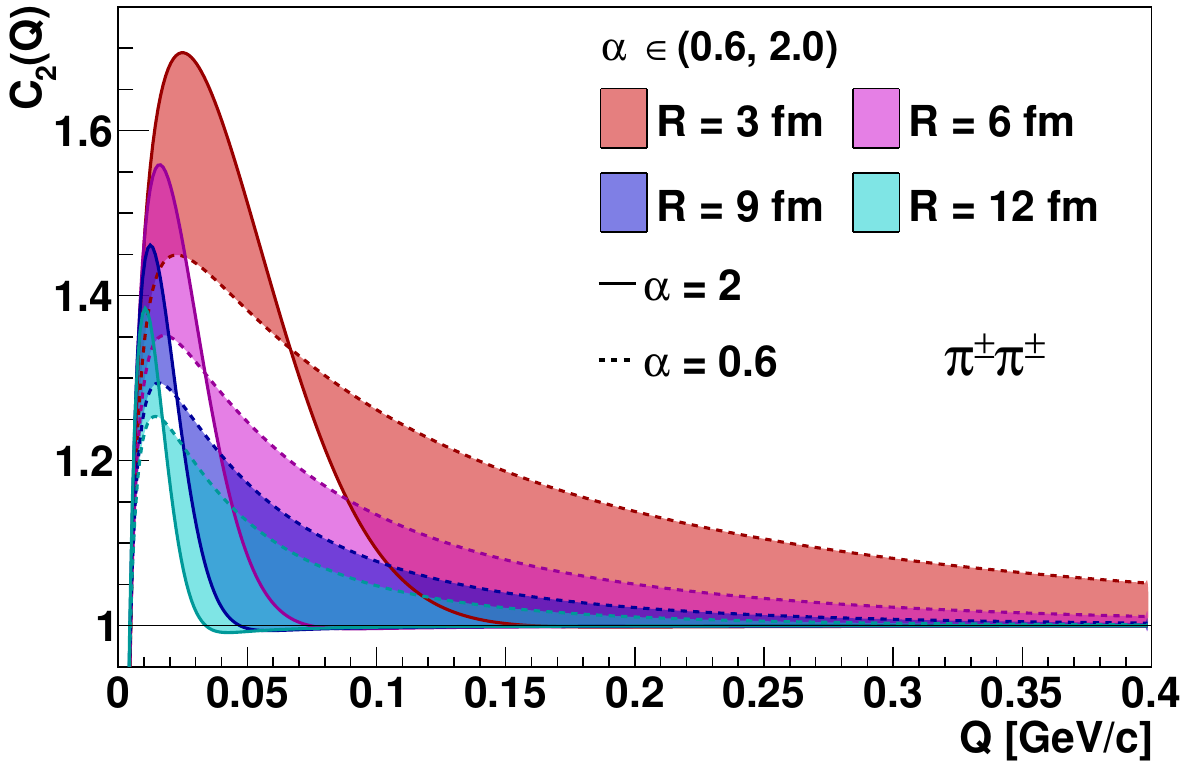}\includegraphics[width=0.5\textwidth]{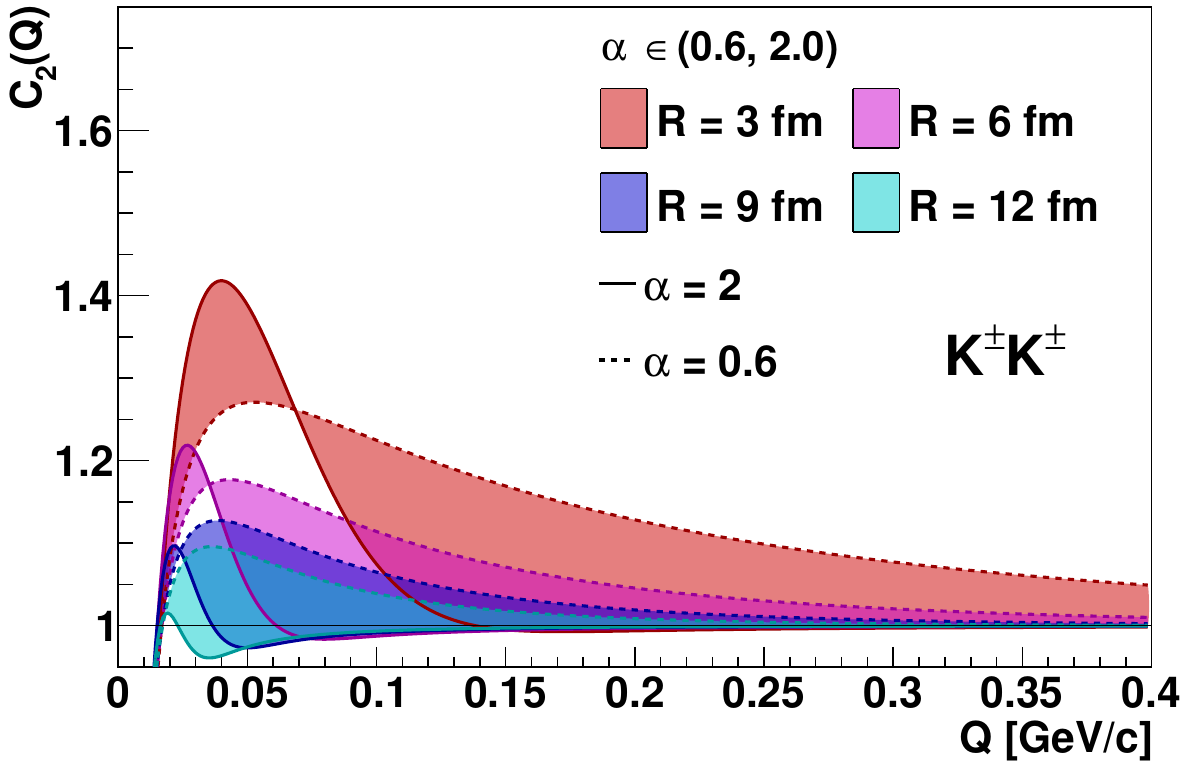}}
    \caption{Example correlation functions for pions (left) and kaons (right), plotted for four different $R$ and two $\alpha$ values. At a given $R$ value, the shape of the correlation function with increasing $\alpha$ values from $\alpha\,{=}\,0.6$ to $\alpha\,{=}\,2$ goes smoothly through the shaded region.}
    \label{fig:compare}
\end{figure}

 
\subsection{Testing the final numerical integral}
\label{ss:integraltest}
As described above in \Eq{e:CwithA1A2}, for the correlation function, two numerical integrals have to be performed. For a reasonable $f_s(q)$ function, such as the one for L\'evy distributions, the integrands there are particularly well-behaving (non-oscillating, smooth) functions. The integrals over $q\in\mathbb R^+_0$ are best transformed to integrals over the $[0,1]$ interval, with a new integration variable $x$ introduced separately on the $q\,{\in}\,[0,2k]$ and the $q\,{\in}\,[2k,\infty]$ intervals as $q\,{=}\,2kx$, and $q\,{=}\,\frac{2k}x$, respectively. The integrands turn out to be well-behaving ,,tame'' functions of this new $x$ variable; they are written up as
\begin{align}
\label{e:A1sexpr_x}
\c A_{1s} = -\frac2\eta\sint{-2pt}{-8pt}01\,\m dx&\,\m{Im}\bigg[\frac{f_s(2kx)\,{-}\,f_s(0)}x\bigg(1{+}\frac{1}x\bigg)^{2i\eta}{}_2F_1\Big(i\eta,1{+}i\eta,1,\rect{x^2}{-}i0\Big)+\\
&\hspace{27pt}+\frac{f_s\big(\frac{2k}x\big)\,{-}\,f_s(0)}x\big(1{+}x\big)^{2i\eta}{}_2F_1\Big(i\eta,1{+}i\eta,1,x^2{-}i0\Big)\bigg],\nonumber\\
\c A_{2s} = -\frac2\eta\sint{-2pt}{-8pt}01\,\m dx&\,\frac{\sin\big(\eta\log\frac{1{+}x}{1{-}x}\big)}{x(x{+}1)}\bigg[\frac{f_s\big(\frac{2k}x\big)\,{-}\,f_s(2k)}{1{-}x}\,{-}\,x^2e^{\pi\eta}\frac{f_s(2kx)\,{-}\,f_s(2k)}{1{-}x}\bigg].
\label{e:A2sexpr_x}
\end{align}
The integrals can then be evaluated using the rectangle method or the trapezoidal rule, but since the individual function evaluations have a high cost in terms of CPU time and in practical applications such as optimization (fit) procedures, the final result has to be calculated several thousands of times at a given $Q$, it is beneficial to search for methods requiring fewer evaluations. We find that the Gauss-Kronrod quadrature formula~\cite{davis2014methods} provides an acceptable solution for this. This method works well in our case because the function to be integrated is changing fast near 0 and 1 but very slowly in the middle. Hence a varying bin width has to be applied, and such naturally arises from the Gauss-Kronrod iteration. 

We utilized the \texttt{boost} C++ library to perform this integral and find that 15 nodes provide a fast converging result. The number of function evaluations and the integral results are shown in Fig.~\ref{fig:integraltest}. Here we chose a large $Q$ value, as the accuracy of the original numerical calculations, when one directly calculates the integral with $|\psi_{\bs k}(\bs r)|^2$, decreases towards large $Q$, and hence accuracy can be well tested. The integral result changes on the order of $10^{-6}$, and even on that scale, it converges when the number of maximum iterations is set to $3-4$. It is also important to note that the typical order of magnitude of the bin-by-bin statistical uncertainties of a correlation function measured in high-energy experiments is significantly higher than the change in the integral values shown in Fig.~\ref{fig:integraltest}. If the tolerance parameter of the quadrature is larger than $10^{-3}$, then the integral result differs slightly more from the plotted results. However, beyond that, if tolerance is at least $10^{-3}$, increasing it further does not modify the integral result. On the other hand, the number of function evaluations increases fast, especially if the tolerance is very small and the number of iterations is large. Hence an optimal solution (analyzing results similar to Fig.~\ref{fig:integraltest} for many $R$, $\alpha$, and $Q$ values) is provided by a tolerance value of $10^{-3}$ and a limit of 3 for the maximal number of iterations. With this, one integral requires up to a few hundred function evaluations, suitable for simultaneously fitting many experimental $C_2(Q)$ datasets.

\begin{figure}
    \centerline{
    \includegraphics[width=\textwidth]{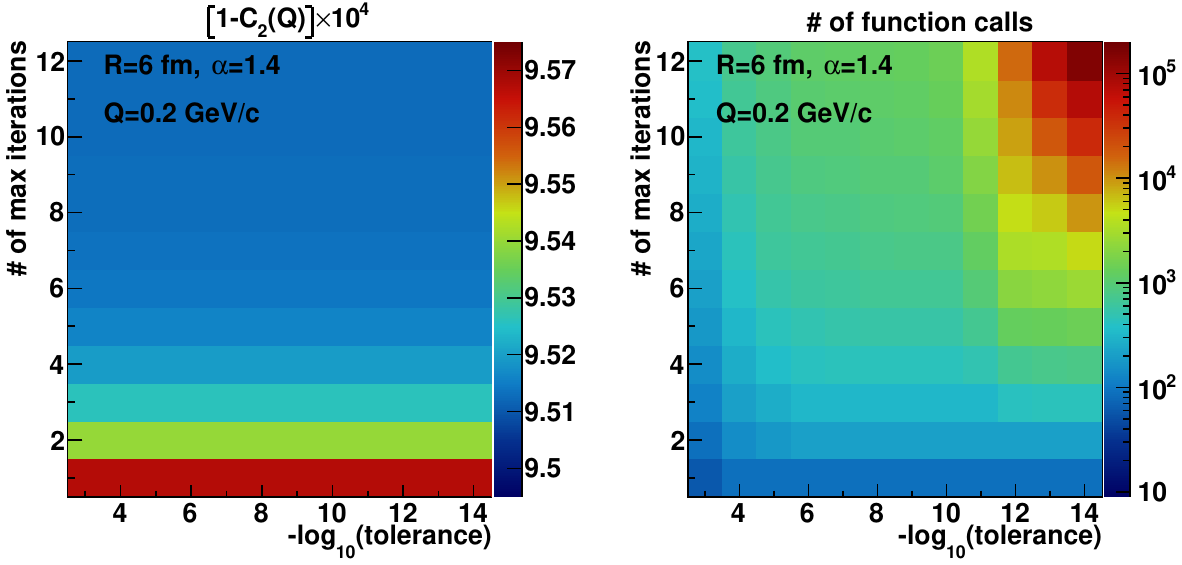}}
    \caption{The result of the integral for $R\,{=}\,6$ fm and $\alpha\,{=}\,1.4$, at $Q\,{=}\,0.2$ GeV$/c$ (left) and the number of required function calls (right). Note that the integral result is represented as $(1-C_2(Q))\cdot10^4$, so the differences are on the order of $10^{-6}$. Both plots are shown as a function of the tolerance and the number of maximal iterations allowed in the Gauss-Kronrod integral.}
    \label{fig:integraltest}
\end{figure}

\subsection{Comparison with the original numerical integration method}\label{ss:comparison}

After carefully testing the reliability of the Gauss-Kronrod numerical integral, we proceeded with comparing the new calculation to a previously utilized integral method based on \Eq{e:CDf0}, and used, e.g., in Ref.~\cite{Kincses:2019rug}. The previous method requires significantly more computational effort to achieve a reasonable precision at small $\alpha$ or large $R$ values. In that case, the values of the correlation function were pre-calculated for various $\alpha$ and $R$ values and saved in a large lookup table; from this table, an interpolation can be used to get the correlation function for any parameter combinations. In the following, let us call our new approach ``wave function Fourier method'' and denote the result with $^{\textnormal{WFF}}C_2(Q)$. The previous method with the pre-calculated lookup table is denoted with $^{\textnormal{table}}C_2(Q)$.

To illustrate the difference between the two methods, we plot ${\Delta C_2(Q) =\; {^{\textnormal{table}}}C_2(Q){-}^{\textnormal{WFF}}C_2(Q)}$ for various $\alpha$ and $R$ values, as shown in Figure~\ref{fig:subtr}. We find that $\Delta C_2(Q)$ has a small dependence on $Q$, generally its magnitude is larger at larger $Q$ values. It is also evident that the difference between the two methods is largest at the smallest $\alpha$ values, as expected. To better illustrate the dependence of $\Delta C_2(Q)$ on $R$ and $\alpha$, in Figure~\ref{fig:subtr_avg} we plot the $Q$-average $\langle\Delta C_2(Q)\rangle$ values. It can be clearly seen that the $\Delta C_2(Q)$ difference between the two methods is smallest at large $\alpha$ and small $R$ values.

\begin{figure}
    \centerline{
    \includegraphics[width=0.5\textwidth]{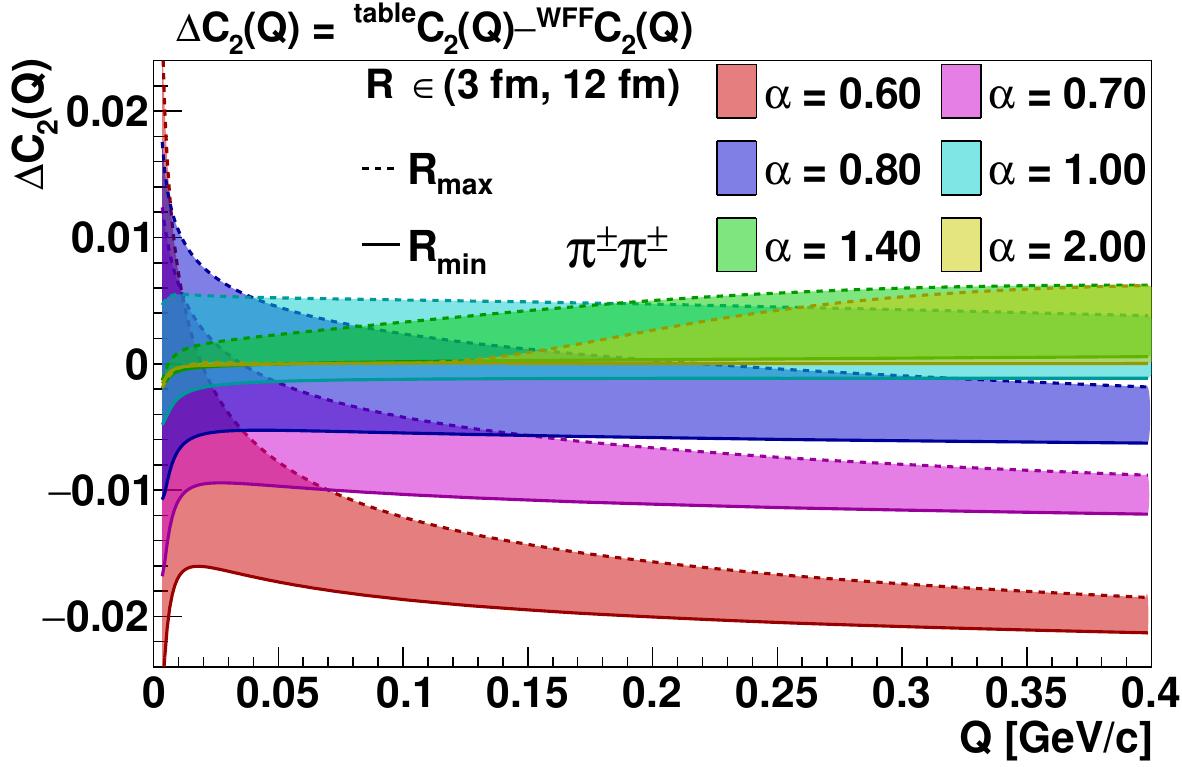}\includegraphics[width=0.5\textwidth]{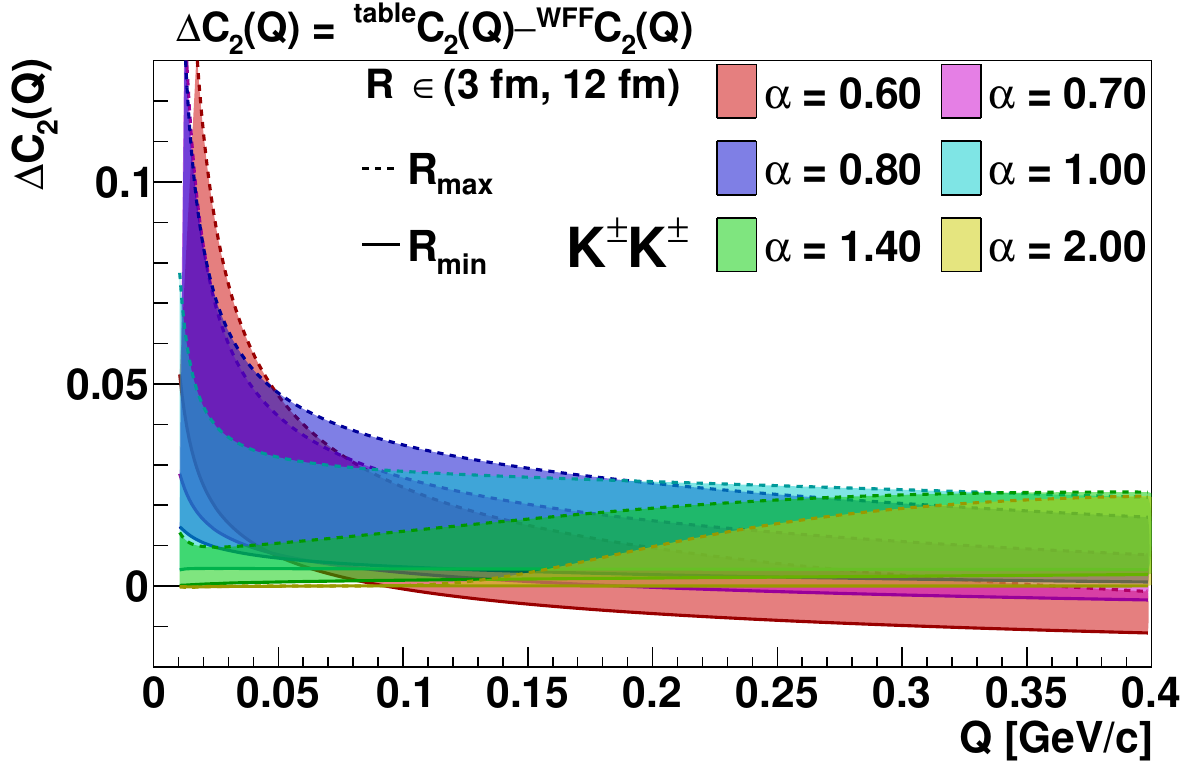}}
    \caption{Difference between the correlation function calculated with a numerical integral method described in Ref.~\cite{Kincses:2019rug} ($^{\textnormal{table}}C_2(Q)$) and the correlation function calculated with the wave-function Fourier method described in the current paper ($^{\textnormal{WFF}}C_2(Q)$). $\Delta C_2(Q)$ is plotted for 6 different $\alpha$ values and two $R$ values, for pions (left) and kaons (right) separately. At a given $\alpha$ value, $\Delta C_2(Q)$ goes smoothly through the shaded region when increasing $R$ values from $R = 3\textnormal{ fm}$ to $R = 12\textnormal{ fm}$.}
    \label{fig:subtr}
\end{figure}

\begin{figure}
    \centerline{
    \includegraphics[width=0.5\textwidth]{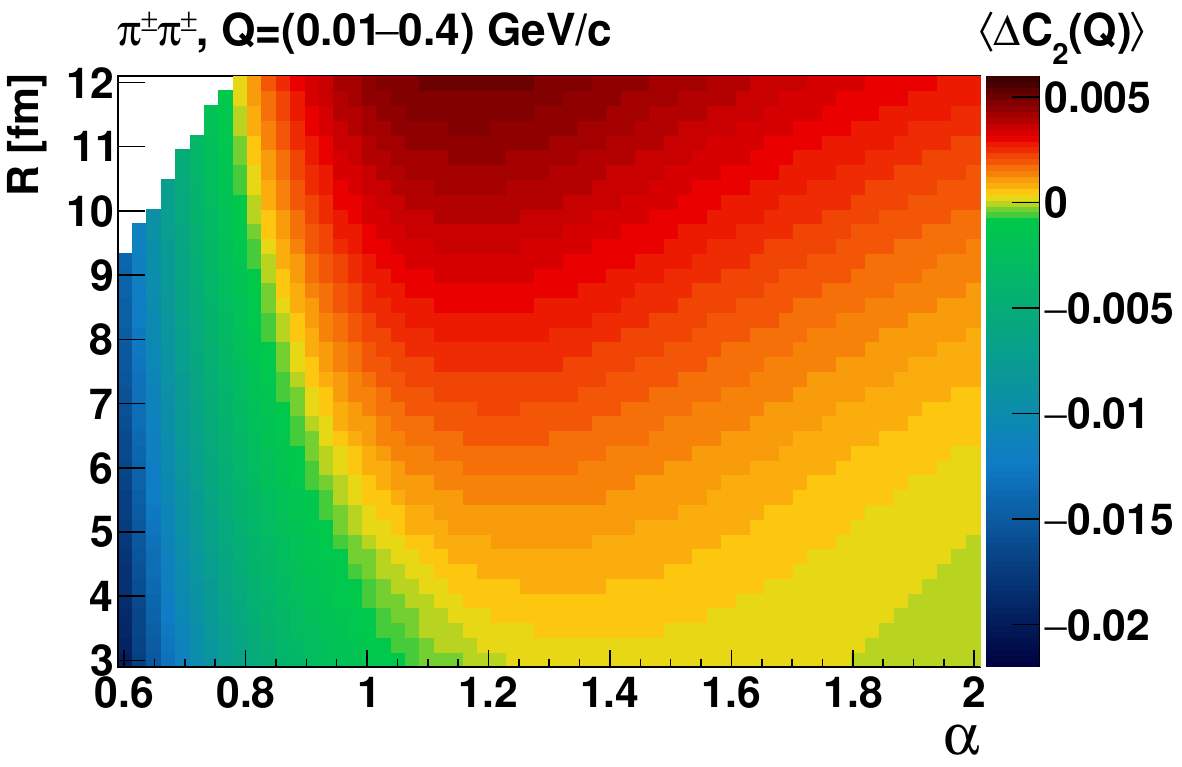}\includegraphics[width=0.5\textwidth]{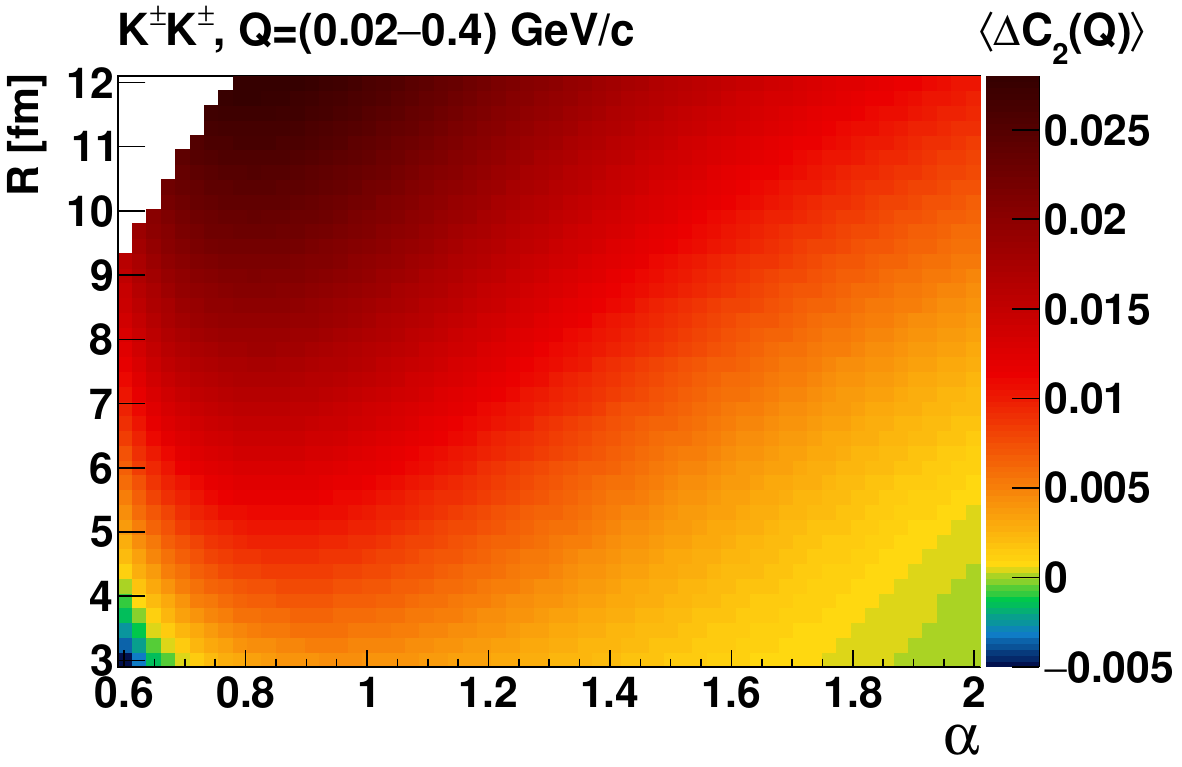}}
    \caption{Relative momentum averaged $\langle\Delta C_2(Q)\rangle$ as a function of $R$ and $\alpha$ for pions (left) and kaons (right).}
    \label{fig:subtr_avg}
\end{figure}
The conclusion of this comparison is that, on one hand, we can be reasonably assured that our new methodology works well; as cautious users, we really strive for numerical verification, even if one is convinced by the flawlessness of the mathematical derivation. On the other hand, our new methodology offers, at a significantly lower numerical cost, a way of calculating the $C_2(Q)$ correlation function more robust than the previous method, especially at large $Q$ values. For this reason, we find that the new method is ready to be used in experimental analyses of measured $C_2(Q)$ correlation functions. 

\section{Summary}\label{s:summary}

In summary, we presented a novel method for calculating Bose-Einstein correlation functions with the Coulomb effect incorporated. Our method can be applied to any particle emitting source function assumptions that are expressed as a Fourier transform. In this way, our method starts directly from the interaction-free correlation function, which is the Fourier transform of the source function. The integrals necessary for our calculational scheme previously have eluded exact calculations, often resulting in experimental analyses using a fixed source size when correcting for the Coulomb effect. With precise data, however, this is no longer a viable method; our approach allows for simple and more exact handling of the Coulomb final state interactions. We demonstrated that for L\'evy-stable sources (including the Gaussian limiting case), the results are close to those obtained with numerical integrals previously. On the other hand, at extreme values of the L\'evy source parameters (small L\'evy exponent $\alpha$, and large L\'evy scale $R$), the new method is more precise and reliable. Furthermore, the new method allows for a direct and fast fitting of correlation functions obtained in experiments. We also published a software package~\cite{CoulCorrLevyIntegral} that is easily applicable and ready to use in experimental analyses.

The natural next step is to extend the methodology to the case when the assumption of spherical symmetry is rid of. In that case, one would have a final formula that is a linear functional of the Fourier transform of the pair source function, $D(\bs r)$, denoted by $f(\bs q)$ in this paper. It turns out that the necessary integrals (similar to but slightly more complicated than \Eqs{e:D1sdef}{e:D2sdef} in this paper) are also readily calculable, and then the necessary limit when the regularizing parameter is removed (denoted as $\lambda\,{\to}\,0$ in this paper) is also manageable. The resulting formulas are, however (owing to the inherently more complex nature of the non-spherically symmetric case) somewhat more complicated, thus their analysis and implementation go beyond the scope of this paper.

\section{Acknowledgements}
This research was supported by the NKFIH grants K-136138 and TKP2021-NKTA-64.

\appendix
\section{Assorted mathematical and calculational details}

In the Appendices, we present the calculations that were left out of the body of the paper. We rely heavily on complex analysis as well as on some intricacies of (Lebesgue) integrability; the latter of which we give a brief summary here. (A more detailed discussion can be found in any of the standard textbooks on mathematical analysis, such as Ref.~\cite{rudin86real}.)

\subsection{Discussion of \Eq{e:CDf1}}\label{ss:app:lebesgue}

The focal point of our method of circumventing the need for a back-and-forth integral transformation of the Fourier transform of the source function (denoted by $f(\bs q)$ in \Eq{e:Dwithf} and from that on) is \Eq{e:CDf1}, the mathematical justification of whose steps is thus essential. The main ingredients are conditions of integrability, \textit{Lebesgue's dominated convergence theorem} (or simply the \textit{Lebesgue theorem}), and \textit{Fubini's theorem}. A summary of these theorems from a practical point of view is:
\begin{itemize}
\item
Concerning Lebesgue integrability, the key is the modulus of a function. A (real-valued, or complex-valued, or any normed space valued) function $F$ is integrable if and only if $|F|$ is integrable. Also, if there is a $G$ function for which $|G(x)|\,{\ge}\,|F(x)|$ for almost all $x$, and $G$ is integrable, then $F$ is also integrable. (Here, we denoted the integration variable by $x$; it can be of any type of real integration variable.) Conversely, if $|G(x)|\,{\ge}\,|F(x)|$, and $F$ is not integrable, then neither is $G$.
\item
Let $F_\lambda(x)$ be integrable functions, with $\lambda$ as a parameter (either a continuous one or a discrete index), and for almost all $x$ let the pointwise limiting function $F(x)\,{:=}\,\lim_\lambda F_\lambda(x)$ exist (for any reasonable type of limit; for example, $\lambda\,{\to}\,0$, or $\lambda\,{\equiv}\,n\,{\to}\,\infty$). \textit{Lebesgue's theorem} states that if there is a ($\lambda$-independent) $G$ function for which for almost any $x$, $|G(x)|\,{\ge}\,|F_\lambda(x)|$, for any $\lambda$, then the integrals of the $F_\lambda$ functions (the $\int F_\lambda$ values) converge, the limiting function $F$ is integrable, and the limit of the integrals is equal to the integral of the limiting function, $\lim_\lambda \int F_\lambda = \int F$, i.e., the limit and the integral are interchangeable.

The key is the existence of the ``dominant'' $G$ function: in the case when there is no such function, none of the statements of the theorem are necessarily true. A simple (counter)example is the approximation of the Dirac delta with functions whose pointwise limit is everywhere zero; in this well-known case, one maybe does not even recognize this peculiarity. However, for less explored cases (like ours in this paper), one has to be careful when interchanging limits and integrals, and then this theorem comes in handy in many cases.
\item
Fubini's theorem concerns multiple integrals, iterated integrals, and the justification of interchange of integrals. The main point is if an $F(x,y)$ function is such that its modulus is integrable in one order as a repeated integral; i.e., if the $\int\m dx\big(\int\m dy\,|F(x,y)|\big)$ integral exists, then $F$ itself is integrable in both orders, and these integrals coincide: $\int\m dx\int\m dy\,F(x,y)\,{=}\,\int\m dy\int\m dx\,F(x,y)$, i.e., the integrals are interchangeable. 
\end{itemize}
The transformation in \Eq{e:CDf1} is again written up, with the intermediate steps slightly more detailed, as
\begin{align}
C_2(\bs Q) &= \sint{-2pt}{-3pt}{}{}\m d^3\bs r\,|\psi_{\bs k}(\bs r)|^2\,D(\bs r)
\stackrel{1.}= \sint{-2pt}{-3pt}{}{}\m d^3\bs r\,\lim_{\lambda\to0}e^{-\lambda r}|\psi_{\bs k}(\bs r)|^2\,D(\bs r)
\stackrel{2.}= \lim_{\lambda\to0}\sint{-2pt}{-3pt}{}{}\m d^3\bs r\,e^{-\lambda r}|\psi_{\bs k}(\bs r)|^2\,D(\bs r)
\stackrel{3.}= \nonumber\\&\qquad \stackrel{3.}=
\lim_{\lambda\to0}\sint{-2pt}{-3pt}{}{}\m d^3\bs r\sint{-2pt}{-3pt}{}{}\frac{\m d^3\bs q}{(2\pi)^3}\,f(\bs q)\,e^{-\lambda r}|\psi_{\bs k}(\bs r)|^2\,e^{i\bs q\bs r}
\stackrel{4.}=
\lim_{\lambda\to0}\sint{-2pt}{-3pt}{}{}\frac{\m d^3\bs q}{(2\pi)^3}\,f(\bs q)\sint{-2pt}{-3pt}{}{}\m d^3\bs r\,e^{-\lambda r}|\psi_{\bs k}(\bs r)|^2\,e^{i\bs q\bs r}.
\end{align}
In Step 1 we inserted the $e^{-\lambda r}$ regularizing factor; its limit being $\lim_{\lambda\to0}e^{-\lambda r}\,{=}\,1$. Because $D(\bs r)$ is integrable, so is it multiplied by $|\psi_{\bs k}(\bs r)|^2$ (which is a bounded function), and thus $|\psi_{\bs k}(\bs r)|^2\cdot|D(\bs r)|$ is a good dominant function (that is, integrable and greater or equal than the integrand, independently of $\lambda$). So by virtue of Lebesgue's theorem, we could interchange the $\lambda\,{\to}\,0$ limit and the $\bs r$-integral in Step 2. In Step 3, we inserted the Fourier integral expression of $D(\bs r)$, careful (yet) about the order of integrals. However, since $|e^{i\bs q\bs r}|\,{=}\,1$, the double integrand here has the modulus that is a product of $|f(\bs q)|$, an integrable $\bs q$-function (as assumed), and an integrable $\bs r$-function, $e^{-\lambda r}|\psi_{\bs k}(\bs r)|^2$. So the repeated integral of the modulus exists, from which it follows (by Fubini's theorem) that we can interchange the original integrals as well; this is Step 4 here.

In the resulting right-hand side, we cannot perform the $\lambda\to0$ limit in the integrands; that would mean the exchange of the original $\bs r$-- and $\bs q$-integrals, which is not possible, as stated around \Eq{e:Cfpsi:wrong}. Instead, we have to calculate the integrals here, and \textit{then} perform the $\lambda\,{\to}\,0$ limit.

\subsection{Calculation of $\c D_{1\lambda s}$ and $\c D_{2\lambda s}$}\label{ss:app:nord}


In this Appendix, we derive the \Eqs{e:D1lambdas}{e:D2lambdas}, the results of the integrals $\c D_{1\lambda s}$ and $\c D_{2\lambda s}$, defined in \Eqs{e:D1sdef}{e:D2sdef}. We utilize Nordsieck's method~\cite{Nordsieck:1954zz}, who applied a similar technique to simplify a very similar integral that occurs in the theory of bremsstrahlung and pair creation. We write up the definitions (\ref{e:D1sdef})--(\ref{e:D2sdef}) again, with a slight change of notation, to express the $\sin(qr)$ function with exponentials:
\begin{align}
&\c D_{1\lambda s}(q) = \rec{2i}\big[\c D_{1\lambda s}^{(+)}(q)-\c D_{1\lambda s}^{(-)}(q)\big],\qquad\qquad
\c D_{2\lambda s}(q) = \rec{2i}\big[\c D_{2\lambda s}^{(+)}(q)-\c D_{2\lambda s}^{(-)}(q)\big],\\
\label{e:app:D1spmdef}
&\qquad\textnormal{where}\quad
\c D_{1\lambda s}^{(\pm)}(q) = \rec q
\sint{-2pt}{-3pt}{}{}\m d^3\bs r\fracd{e^{-\lambda r}}re^{\pm iqr}M\big(1{+}i\eta,1,-i(kr{+}\bs k\bs r)\big)M\big(1{-}i\eta,1,i(kr{+}\bs k\bs r)\big),\\
&\;\;\qquad\textnormal{and}\quad\;\c D_{2\lambda s}^{(\pm)}(q) = \rec q
\sint{-2pt}{-3pt}{}{}\m d^3\bs r\fracd{e^{-\lambda r}}re^{\pm iqr}M\big(1{+}i\eta,1,-i(kr{-}\bs k\bs r)\big)M\big(1{-}i\eta,1,i(kr{+}\bs k\bs r)\big).
\label{e:app:D2spmdef}
\end{align}
We use the following complex contour integral representation of the confluent hypergeometric function:
\begin{align}
M(a,1,z) = \rec{2\pi i}\soint{-2pt}{-16pt}{}{(0+,1+)}{}\m dt\,e^{tz}\rect t\big(1{-}\rect t\big)^{-a},
\label{e:confhyp:contour}
\end{align}
where the path is any closed curve that encircles the real line segment $[0,1]$ once counterclockwise on the complex $t$ plane; this segment (a branch cut) is the only set on the $t$ plane where the integrand is not analytic.

Inserting twice this expression (with the integration variables denoted by $t$ and $u$, whose paths are such as just specified) into \Eqs{e:app:D1spmdef}{e:app:D2spmdef} for the two confluent hypergeometric functions in each, we get
\begin{align}
\label{e:D1scontour0}
&\c D_{1\lambda s}^{(\pm)}(q) = -\rec{4\pi^2q}\sint{-3pt}{-2pt}{}{}\frac{\m d^3\bs r}re^{-\lambda r\pm iqr}\soint{-2pt}{-3pt}{}{}\frac{\m du}u\soint{-2pt}{-3pt}{}{}\frac{\m dt}t
\big(1{-}\rect t\big)^{-1-i\eta}\big(1{-}\rect u\big)^{-1+i\eta}e^{-i(t-u)(kr+\bs k\bs r)},\\
&\c D_{2\lambda s}^{(\pm)}(q) = -\rec{4\pi^2q}\sint{-3pt}{-2pt}{}{}\frac{\m d^3\bs r}re^{-\lambda r\pm iqr}\soint{-2pt}{-3pt}{}{}\frac{\m du}u\soint{-2pt}{-3pt}{}{}\frac{\m dt}t
\big(1{-}\rect t\big)^{-1-i\eta}\big(1{-}\rect u\big)^{-1+i\eta}e^{-i(t-u)kr+i(t+u)\bs k\bs r}.
\label{e:D2scontour0}
\end{align}
We would like to exchange the order of the integral over $\bs r$ and the contour integrals because for the $\bs r$-integral, we could then use the following auxiliary formula, valid for any $\beta\,{\in}\,\mathbb C$ complex number and $\bs B\,{\equiv}\,(B_x,B_y,B_z)$ three-vector with any \textit{complex} components:
\begin{align}
\sint{0pt}{-2pt}{}{}\frac{\m d^3\bs r}re^{-\beta r+\bs B\bs r}=\frac{4\pi}{\beta^2{-}\bs B^2},\quad\textnormal{if}\quad \m{Re}\,\beta \,{>}\,|\m{Re}\,\bs B|.
\label{e:spaceint1}
\end{align}
Here $\bs B^2\,{=}\,B_x^2{+}B_y^2{+}B_z^2$ even for complex $B_x$, $B_y$, and $B_z$ (in particular, \textit{without} complex conjugation), and the real part of the $\bs B$ vector is $\m{Re}\,\bs B\,{:=}\,(\m{Re}\,B_x, \m{Re}\,B_y, \m{Re}\,B_z)$, thus $|\m{Re}\,\bs B|\,{=}\,\sqrt{(\m{Re}\,B_x)^2\,{+}\,(\m{Re}\,B_y)^2\,{+}\,(\m{Re}\,B_z)^2}$. The condition written up in \Eq{e:spaceint1} is necessary and sufficient for the integral to exist. This is because a function is (Lebesgue) integrable if and only if its modulus is integrable, and $\big|\rec re^{-\beta r+\bs B\bs r}\big|\,{=}\,\rec re^{-\m{Re}\,\beta\cdot r+(\m{Re}\bs B)\bs r}$, which is integrable if and only if the multiplier of $r$ in the exponent, $\m{Re}\,\beta$, is strictly bigger than the length of the vector there, $|\m{Re}\,\bs B|$. For real $\beta$, $B_x$, $B_y$, $B_z$, the stated result is elementary, and because both the integral itself and the stated result are analytic functions of $\beta$, $B_x$, $B_y$, $B_z$ (provided the integral exists), the result is valid for all the allowed complex values, as stated.

If we can exchange the $\bs r$-integral and the contour integrals in \Eqs{e:D1scontour0}{e:D2scontour0}, the we would get
\begin{align}
\label{e:D1scontour1}
&\c D_{1\lambda s}^{(\pm)}(q) \stackrel{?}= -\rec{4\pi^2q}\soint{-2pt}{-3pt}{}{}\frac{\m du}u\soint{-2pt}{-3pt}{}{}\frac{\m dt}t
\big(1{-}\rect t\big)^{-1-i\eta}\big(1{-}\rect u\big)^{-1+i\eta} \sint{-3pt}{-2pt}{}{}\frac{\m d^3\bs r}re^{-\lambda r\pm iqr-i(t-u)(kr+\bs k\bs r)},\\
&\c D_{2\lambda s}^{(\pm)}(q) \stackrel{?}= -\rec{4\pi^2q}\soint{-2pt}{-3pt}{}{}\frac{\m du}u\soint{-2pt}{-3pt}{}{}\frac{\m dt}t
\big(1{-}\rect t\big)^{-1-i\eta}\big(1{-}\rect u\big)^{-1+i\eta}\sint{-3pt}{-2pt}{}{}\frac{\m d^3\bs r}re^{-\lambda r\pm iqr-i(t-u)kr+i(t+u)\bs k\bs r}.
\label{e:D2scontour1}
\end{align}
According to \textit{Fubini's theorem}, the interchange is justified if the \textit{modulus} of the integrand is integrable in either order (provided that one has written up the contour integrals in a parametrized way, with integrals taken over line segment of the parameter). In our case, the moduli of the complex powers of $t$ and $u$ do not cause concern: since $u$ and $t$ run at a positive distance from the branch cut, these factors are bounded and thus do not spoil integrability. The factor of the modulus of the integrand that indeed has a role is that of the $\bs r$-dependent part. In the case of $\c D_{1\lambda s}$, \Eq{e:D1scontour1}, the modulus of the part of the integrand that is of interest is 
\begin{align*}
\big|e^{-\lambda r\pm iqr-i(t-u)(kr+\bs k\bs r)}\big| = e^{-[\lambda-k\m{Im}(t-u)]r+\m{Im}(t-u)\bs k\bs r},
\end{align*}
which is integrable over $\bs r$ if and only if for the following holds (by virtue of the condition stated in \Eq{e:spaceint1}:
\begin{align}
\frac\lambda k-\m{Im}(t{-}u)\,{>}\,|\m{Im}(t{-}u)| \qquad\Leftrightarrow\qquad \m{Im}(t{-}u)\,{<}\,\frac\lambda{2k}\qquad\textnormal{(for $\c D_{1\lambda s}$).}
\label{e:cond:D1s}
\end{align}
We arrived at the simpler condition by sorting through all the different possibilities for $\m{Im}\,u$ and $\m{Im}\,t$.
Similarly, in the case of $\c D_{2\lambda s}$, \Eq{e:D2scontour1}, we have 
\begin{align*}
\big|e^{-\lambda r\pm iqr-i(t-u)kr+i(t+u)\bs k\bs r}\big| = e^{-[\lambda-k\m{Im}(t-u)]r-\m{Im}(t+u)\bs k\bs r},
\end{align*}
which is integrable over $\bs r$ if and only if
\begin{align}
\frac\lambda k-\m{Im}(t{-}u)\,{>}\,|\m{Im}(t{+}u)| \qquad\Leftrightarrow\qquad
\m{Im}\,u\,{>}\,{-}\frac\lambda{2k}\quad\textnormal{and}\quad\m{Im}\,t\,{<}\,\frac\lambda{2k}\qquad\textnormal{(for $\c D_{2\lambda s}$).}
\label{e:cond:D2s}
\end{align}
If these conditions hold for all possible $u$ and $t$, then the results of these $\bs r$-integrals are continuous bounded functions of $u$ and $t$, so their contour integrals exist. So the condition for the integrals over $\bs r$ and $u,t$ to be interchangeable is that the inequalities stated in \Eqs{e:cond:D1s}{e:cond:D2s} hold for any $u$ and $t$ on their integration paths. Thus if we require these additional constraints on the $u$-- and $t$-paths, we can interchange the integrals, and utilizing \Eq{e:spaceint1}, we get the following from \Eqs{e:D1scontour1}{e:D2scontour1}:
\begin{align}
\label{e:D1scontour2}
&\c D_{1\lambda s}^{(\pm)}(q) = -\rec{\pi q}\rec{\lambda{\mp}iq}\soint{-2pt}{-3pt}{}{}\frac{\m dt}t\big(1{-}\rect t\big)^{-1-i\eta}\soint{-2pt}{-3pt}{}{}\frac{\m du}u
\big(1{-}\rect u\big)^{-1+i\eta}\rec{\lambda{\mp}iq{+}2ik(t{-}u)},\\
&\c D_{2\lambda s}^{(\pm)}(q) = -\rec{\pi q}\soint{-2pt}{-3pt}{}{}\frac{\m dt}t\big(1{-}\rect t\big)^{-1-i\eta}\rec{2kt{-}i(\lambda{\mp}iq)}
\soint{-2pt}{-3pt}{}{}\frac{\m du}u\big(1{-}\rect u\big)^{-1+i\eta}\rec{2ku{+}i(\lambda{\mp}iq)},
\label{e:D2scontour2}
\end{align}
where the paths of the $u$ and $t$ variables run as specified: they both encircle the $[0,1]$ branch cut and obey at all points the conditions (\ref{e:cond:D1s}) in case of $\c D_{1\lambda s}$, and (\ref{e:cond:D2s}) in case of $\c D_{2\lambda s}$, respectively. (For any finite $\lambda$ and $k$, it is indeed possible to specify the integration paths in this way.) The case of $\c D_{2\lambda s}$ is simpler because the integrand factorizes in $u$ and a $t$; we have written \Eq{e:D2scontour2} in a way that highlights this.

The next step is to perform the $u$-integral at a fixed $t$. As a function of $u$, the integrand in both cases is analytic everywhere except for the branch cut along $u\,{\in}\,[0,1]$ which is encircled by the path, as well as for a simple pole, denoted in case of $\c D_{1\lambda s}$ and $\c D_{2\lambda s}$ by $u_{1s}(t)$ and $u_{2s}(t)$, respectively: 
\begin{align}
\label{e:u1su2s}
u_{1s}(t) = t\mp\frac q{2k}{-}\frac{i\lambda}{2k},\qquad
u_{2s}(t) = -\frac{i\lambda}{2k}{\mp}\frac q{2k}.
\end{align}
We see that $\m{Im}(t{-}u_{1s})\,{=}\,\frac\lambda{2k}$ and $\m{Im}(u_{2s})\,{=}\,{-}\frac\lambda{2k}$: this means that (for any $t$ that is in its allowed domain) $u_{1s}$ and $u_{2s}$ do not satisfy the conditions in \Eqs{e:cond:D1s}{e:cond:D2s}. From this, one concludes that $u_{1s}$ and $u_{2s}$ lies outside of the integration contour on the $u$ plane: were it otherwise, a narrower integration path chosen for $u$ could cross $u_{1s}$ and $u_{2s}$, but this is impossible because then $u_{1s}$ and $u_{2s}$ could not violate the conditions.

One can then expand the $u$-integration contour to infinity: the integrand decreases rapidly enough (as $\sim\rec{u^2}$ in both cases), so the non-vanishing contribution comes from the simple poles $u_{1s}$ and $u_{2s}$, so from the residue theorem we have (with an extra minus sign from the negative sense of the paths encircling the poles)
\begin{align}
\label{e:D1scontour3}
&\c D_{1\lambda s}^{(\pm)}(q) = -\rec q\soint{-3pt}{-3pt}{}{}\frac{\m dt}t\frac{\big(1{-}\rect t\big)^{-1-i\eta}}{k(\lambda{\mp}iq)}\rec{u_{1s}}
\bigg(1{-}\rec{u_{1s}}\bigg)^{-1+i\eta},\\
&\c D_{2\lambda s}^{(\pm)}(q) = -\rec q\frac2{\lambda{\mp}iq}\bigg(\frac{\lambda{\mp}iq{-}2ik}{\lambda{\mp}iq}\bigg)^{-1+i\eta}
\soint{-3pt}{-3pt}{}{}\frac{\m dt}t\frac{\big(1{-}\rect t\big)^{-1-i\eta}}{2kt{-}i(\lambda{\mp}iq)},
\label{e:D2scontour3}
\end{align}
where in case of $\c D_{1\lambda s}$, $u_{1s}(t)$ is as in \Eq{e:u1su2s}, while in case of $\c D_{2\lambda s}$ we inserted $u_{2s}$ from \Eq{e:u1su2s} right away.%
\footnote{
Perhaps it should be stressed that expanding the $u$-contour here means no conflict with our earlier requirement on the $u$-path, \Eqs{e:cond:D1s}{e:cond:D2s}. At that stage, that restriction was needed in order to interchange the integrals; after that (and the $\bs r$-integral performed), the resulting integrand is analytic in $u$, so we may do this expansion in this stage.
}

In \Eq{e:D2scontour3}, we can use this same method for the $t$-integral. The integrand has a simple pole in $t$ at $t_{2s}\,{=}\,\frac{i\lambda}{2k}\,{\pm}\,\frac q{2k}$, and the same reasoning as for $u_{2s}$ above yields that because $\m{Im}\,t_{2s}\,{=}\,\frac\lambda{2k}$, $t_{2s}$ does not satisfy the condition in \Eq{e:cond:D2s}, so it lies outside of the integration contour. Thus an expansion of the contour to infinity yields (again because of the $\sim\rec{t^2}$ decrease) just the contribution from the pole at $t_{2s}$. From this, we get
\begin{align}
\c D_{2\lambda s}^{(\pm)}(q) = \frac{4\pi}q\rec{\lambda{\mp}iq}\bigg(\frac{\lambda{\mp}iq{-}2ik}{\lambda{\mp}iq}\bigg)^{-1+i\eta}
\rec{\lambda{\mp}iq}\bigg(\frac{\lambda{\mp}iq{+}2ik}{\lambda{\mp}iq}\bigg)^{-1-i\eta}.
\end{align}
This leads to the result for $\c D_{2\lambda s}$, stated in \Eq{e:D2lambdas} in the main text, after some simplification.

In the case of $\c D_{1\lambda s}$, the remaining integrand in \Eq{e:D1scontour3} does not have a simple pole as a function of $t$ (besides the branch cut on $[0,1]$), but another branch cut on a straight line segment between $t\,{=}\,{\pm}\frac q{2k}{+}i\frac\lambda{2k}$ and $t\,{=}\,1{\pm}\frac q{2k}{+}i\frac\lambda{2k}$ (i.e., where $u_{1s}(t)\,{=}\,0$ and $u_{1s}(t)\,{=}\,1$). This branch cut, as its imaginary part, $\frac\lambda{2k}$ is in violation of condition (\ref{e:cond:D1s}), lies outside the $t$ integration contour. In this case, we can transform the integral in \Eq{e:D1scontour3} into a form that yields an ordinary hypergeometric function. To this end, we substitute $t = \rec s$, with which we arrive at 
\begin{align}
&\c D_{1\lambda s}^{(\pm)}(q) = N\soint{-3pt}{-2pt}{}{}\m ds\frac{(1{-}s)^{-1-i\eta}}{s{-}A}
\bigg[\beta\frac{s{-}B}{s{-}A}\bigg]^{-1+i\eta},\qquad\textnormal{where we used the notations}\\
&\qquad N\,{=}\,\frac{-2i}{q(\lambda{\mp}iq)^2},\qquad A\,{:=}\,\frac{2k}{i\lambda{\pm}q}, \qquad
B\,{:=}\,\frac{2k}{i\lambda{\pm}q{+}2k},\qquad \beta\,{:=}\,1{+}\frac{2k}{i\lambda{\pm}q},
\end{align}
and the integration contour on the $s$ plane does not intersect the $[1,\infty]$ half line (the image of the original $t\,{\in}\,[0,1]$ cut), but encircles $s{=}0$ (the image of $t{=}\infty$) and the cut between $s{=}A$ and $s{=}B$ (which is a circle segment; the image of the outlying branch cut on the $t$ plane). 

We need now a linear transformation of the integration variable $s\,{\to}\,\tau$, so that the three singular points $s\,{=}\,1$, $s\,{=}\,A$ and $s\,{=}\,B$ transform into $\tau\,{=}\,0$, $\tau\,{=}\,1$ and $\tau\,{=}\,x$ with some $x$. This $x$ will then be the variable of the resulting hypergeometric function; this is motivated by the integral representation
\begin{align}
\label{e:hypintrepresent}
{}_2F_1\big(a,b,c,z\big) = \frac{\Gamma(c)\Gamma(b{-}c{+}1)}{\Gamma(b)}\rec{2\pi i}\sint{-2pt}{-30pt}{\tau=-\infty}{(0+,z+)}
\m d\tau\,\tau^{a-c}(1{-}\tau)^{c-b-1}(\tau{-}x)^{-a} ,
\end{align}
with the $\tau$ integration path coming from $\m{Re}\,\tau\,{=}\,{-}\infty$, going back there, while encircling the half line branch cuts $\tau\,{\in}\,\mathbb R_0^-$ and $\tau{-}x\,{\in}\,\mathbb R_0^-$, but not intersecting $\tau\,{\in}\,[1,\infty]$, the third branch cut of the integrand.%
\footnote{
\Eq{e:hypintrepresent} is well known in the theory of hypergeometric functions; its equivalence to \Eq{e:hypdefpowerseries} for $|x|\,{<}\,1$ can be verified by a power series expansion of the present integrand in $x$ and using similar representations of the products of gamma functions. \Eq{e:hypintrepresent} also gives the analytic continuation of $_2F_1$, since it defines an analytic function of $x$ for any $x\,{\in}\,\mathbb C$, with a branch cut on $x\,{\in}\,[1,\infty]$.
}

The most convenient $s\,{\to}\,\tau$ transformation (among the many possible ones) turns out to be the following:
\begin{align}
s\,{=}\,B\,{+}\,(1{-}B)\tau;
\end{align}
this takes $s{=}B$ into $\tau{=}0$, $s{=}1$ into $\tau{=}1$, and $s{=}A$ into $\tau{=}\frac{A-B}{1-B}$, and we get
\begin{align}
\c D_{1\lambda s} = N\soint{-3pt}{-24pt}{}{(0+,x+)}\m d\tau\,((1{-}B)(1{-}\tau))^{-1-i\eta}\rec{\tau{-}x}\bigg(\frac{\beta\tau}{\tau{-}x}\bigg)^{-1+i\eta},
\qquad\textnormal{with}\quad x=\frac{A{-}B}{1{-}B} = \frac{4k^2}{(q{\pm}i\lambda)^2}.
\label{e:nord:int2}
\end{align}
Now in order to arrive at a form resembling the integral representation (\ref{e:hypintrepresent}), we have to simplify the complex powers. This is not entirely trivial because owing to the branch cut of power functions, for complex numbers, the identity $(ab)^c\,{=}\,a^cb^c$ holds only if $\m{arg}\,a\,{+}\,\m{arg}\,b\,{=}\,\m{arg}(ab)$, i.e., if ${-}\pi\,{<}\,\m{arg}\,a\,{+}\,\m{arg}\,b\,{\le}\pi$. If the displayed quantities are functions of some other complex variable $t$, it is very cumbersome to verify this condition directly. For line integrals, a workaround is that if the identity $(ab)^c\,{=}\,a^cb^c$ is verified in a small neighborhood of some point $t$ on the path, and both sides are analytic in $t$ in the vicinity of the path (practically: the path does not intersect branch cuts of either form on the $t$ plane), then owing to the rigidity of analytic functions, the substitution is justified along the whole integration path.

In the case of \Eq{e:nord:int2}, the steps are as follows. We do not write up all the details about these transformations here; the interested reader is advised to directly check that these steps are indeed justified. 
\begin{enumerate}
\item 
First we do the $((1{-}B)(1{-}\tau))^{-1-i\eta}\,{=}\,(1{-}B)^{-1-i\eta}(1{-}\tau)^{-1-i\eta}$ substitution. For this, we must verify that the integration path on the $\tau$ plane can be specified (and we indeed do so) in a way that it does not intersect the branch cuts of either of these forms (half lines, but in different directions).
\item 
Next, we substitute $\big(\frac{\beta\tau}{\tau{-}x}\big)^{-1+i\eta}$ with $\beta^{-1+i\eta}\big(\frac\tau{\tau{-}x}\big)^{-1+i\eta}$. For this, we must verify that we can take the path so that it does not intersect the branch cuts of either form (a circle segment and a line segment, respectively); for this, the key element is that none of these cuts intersect with the other branch cuts of the integrand.
\item
Now we open up the contour so that instead of being a closed one encircling $0$ and $x$, it comes from $\m{Re}\,\tau\,{=}\,{-}\infty$ and returns there. This can be done because, owing to the exponents, the integrand decreases fast enough (as $\sim\rec{\tau^2}$) at $|\tau|\,{\to}\,\infty$.
\item
Finally, we can write $\big(\frac\tau{\tau{-}x}\big)^{-1+i\eta}\,{=}\,\tau^{-1+i\eta}(\tau{-}x)^{1-i\eta}$; besides $\tau\in[1,\infty]$, this latter form has branch cuts that are straight half lines from $\m{Re}\,\tau\,{=}\,{-}\infty$ to $0$ and $x$, respectively. 
\end{enumerate}
Finally, we thus arrive at the following expression, where we can utilize the (\ref{e:hypintrepresent}) integral representation: 
\begin{align}
\c D_{1\lambda s} = N\frac{\beta^{-1+i\eta}}{(1{-}B)^{1+i\eta}}\sint{-3pt}{-24pt}{\tau=-\infty}{(0+,x+)}\m d\tau\,\tau^{-1+i\eta}(1{-}\tau)^{-1-i\eta}(\tau{-}x)^{-i\eta}
= 2\pi i\,N\frac{\beta^{-1+i\eta}}{(1{-}B)^{1+i\eta}}{}_2F_1(i\eta,1{+}i\eta,1,x),
\end{align}
and from this we get the result given for $\c D_{1\lambda s}$ in the body of the paper, \Eq{e:D1lambdas}, by substituting $N$, $\beta$, $B$, and $x$, and finally combining together $\c D_{1\lambda s}^{(+)}$ and $\c D_{1\lambda s}^{(-)}$ again, according to \Eq{e:app:D1spmdef}.

\subsection{Derivation of the final formula, \Eq{e:CwithA1A2}}\label{ss:app:finallimit}

In this Appendix, we derive \Eq{e:CwithA1A2} along the expression of the quantities denoted by $\c A_{1s}$ and $\c A_{2s}$ in the body of the paper, \Eqs{e:A1sexpr}{e:A2sexpr}. The starting point is \Eq{e:CQspherD1D2}, the expression of $C_2(Q)$ with $\c D_{1\lambda s}$ and $\c D_{2\lambda s}$ as a $\lambda\,{\to}\,0$ limit. We write this up again but first with the integral denoted as a three-dimensional $\bs q$-integral, and perform the first crucial step as we ``separate'' the $f_s(0)$ and $f_s(2k)$ values. This idea comes from observing that the values of $f_s(q)$ at $q\,{=}\,0$ and $q\,{=}\,2k$ play a special role in the interaction-free case as $C_2^{(0)}(Q)\,{=}\,f_s(0){+}f_s(2k)$, moreover, $f_s(0)$ and $f_s(2k)$ enter by virtue of the ``plain'' and the ``cross'' terms from the modulus square of the wave function, just as $\c D_{1\lambda s}$ and $\c D_{2\lambda s}$ in the interacting case. We thus write

\begin{align}
C_2(Q) &= \frac{|\c N|^2}{8\pi^3}\lim_{\lambda\to0}\sint{-2pt}{-3pt}{}{}\m d^3\bs q\,f_s(q)\big[\c D_{1\lambda s}(q)\,{+}\,\c D_{2\lambda s}(q)\big] = \nonumber\\
&= \frac{|\c N|^2}{8\pi^3}f_s(0)\lim_{\lambda\to0}\sint{-2pt}{-3pt}{}{}\m d^3\bs q\,\c D_{1\lambda s}(q)+
\frac{|\c N|^2}{8\pi^3}\lim_{\lambda\to0}\sint{-2pt}{-3pt}{}{}\m d^3\bs q\,\big[f_s(q){-}f_s(0)\big]\c D_{1\lambda s}(q) + \nonumber\\
&\quad + \frac{|\c N|^2}{8\pi^3}f_s(2k)\lim_{\lambda\to0}\sint{-2pt}{-3pt}{}{}\m d^3\bs q\,\c D_{2\lambda s}(q)+
\frac{|\c N|^2}{8\pi^3}\lim_{\lambda\to0}\sint{-2pt}{-3pt}{}{}\m d^3\bs q\,\big[f_s(q){-}f_s(2k)\big]\c D_{2\lambda s}(q).
\end{align}
Now for $\lambda{>}0$, the integrals of $\c D_{1\lambda s}$ and $\c D_{2\lambda s}$, denoted temporarily by $I_1$ and $I_2$, do exist:
\begin{align}
I_1:=\sint{-2pt}{-3pt}{}{}\m d^3\bs q\,\c D_{1\lambda s}(\bs q) = 4\pi\sint{-2pt}{-10pt}0\infty\m dq\,q^2D_{1\lambda s}(q),\qquad
I_2:=\sint{-2pt}{-3pt}{}{}\m d^3\bs q\,\c D_{2\lambda s}(\bs q) = 4\pi\sint{-2pt}{-10pt}0\infty\m dq\,q^2D_{2\lambda s}(q), 
\end{align}
which can be directly seen from their expressions, \Eqs{e:D1lambdas}{e:D2lambdas}: $q^2\c D_{1\lambda s}(q)$ and $q^2\c D_{2\lambda s}(q)$ are continuous and bounded on every compact $[0,q_{\m{max}}]$ interval, and they decrease as $\sim\rec{q^2}$, owing to the taking of the imaginary part in them; recall that ${}_2F_1(a,b,c,x)\,{=}\,1\,{+}\,\frac{ab}c x\,{+}\,\ldots$ as $x\,{\to}\,0$. So the $I_1$ and $I_2$ integrals are given by the value of the Fourier transforms of $\c D_{1\lambda s}$ and $\c D_{2\lambda s}$ at $\bs r\,{=}\,0$ (where $\bs r$ is the variable of their Fourier transforms).%
\footnote{
Recall that the Fourier transform of an integrable function is not necessarily integrable; if it is (which is so in our case; but we had to explicitly check this), then its (inverse) Fourier transform is indeed given by the usual Fourier integral.
}
But we actually calculated $\c D_{1\lambda s}$ and $\c D_{2\lambda s}$ as Fourier transforms, so we just need to evaluate the integrands of the defining (\ref{e:D1sdef})--(\ref{e:D2sdef}) integrals at $\bs r\,{=}\,0$, from which, knowing that $M(a,b,x{=}0)\,{=}\,1$, we have $I_1\,{=}\,I_2\,{=}\,8\pi^3$, so
\begin{align}
&C_2(Q) = |\c N|^2\bigg[f_s(0) + f_s(2k) + \lim_{\lambda\to0}\sint{-2pt}{-10pt}0\infty\m dq\,\frac{q^2\c D_{1\lambda s(q)}}{2\pi^2}\big[f_s(q){-}f_s(0)\big]
+ \lim_{\lambda\to0}\sint{-2pt}{-10pt}0\infty\m dq\,\frac{q^2\c D_{2\lambda s}(q)}{2\pi^2}\big[f_s(q){-}f_s(2k)\big]\bigg].
\end{align}
Our statement is now that these remaining limits result in $\c A_{1s}$ and $\c A_{2s}$ as written up in \Eqs{e:A1sexpr}{e:A2sexpr}. Substituting $\c D_{1\lambda s}$ and $\c D_{2\lambda s}$ from \Eqs{e:D1lambdas}{e:D2lambdas}, and using the convenient $\c F_+(x)\,{\equiv}\,{}_2F_1(i\eta,1{+}i\eta,1,x)$ notation, we thus have to prove that 
\begin{align}
\label{e:app:finallimitD1s}
&\lim_{\lambda\to0}\sint{-2pt}{-10pt}0\infty\m dq\,q\big[f_s(q){-}f_s(0)\big]\m{Im}\bigg[\frac{\big(1{+}\fract{2k}{q{+}i\lambda}\big)^{2i\eta}}{(\lambda{-}iq)^2}\c F_+\Big(\fract{4k^2}{(q{+}i\lambda)^2}\Big)\bigg] 
= \sint{-2pt}{-10pt}0\infty\m dq\,\frac{f_s(q){-}f_s(0)}q\m{Im}\Big[\big(1{+}\fract{2k}q\big)^{2i\eta}{}\c F_+\Big(\fract{4k^2}{q^2}{-}i0\Big)\bigg],\\
&\lim_{\lambda\to0}\sint{-2pt}{-10pt}0\infty\m dq\,q\big[f_s(q){-}f_s(2k)\big]\m{Im}\frac{(\lambda{-}iq{-}2ik)^{i\eta}(\lambda{-}iq{+}2ik)^{-i\eta}}{(\lambda{-}iq)^2{+}4k^2} 
= -\sint{-2pt}{-10pt}0\infty\m dq\,\frac{f_s(q){-}f_s(2k)}{q{-}2k}\frac q{q{+}2k}\m{Im}\frac{(q{+}2k)^{i\eta}}{(q{-}2k{+}i0)^{i\eta}} .
\label{e:app:finallimitD2s}
\end{align}
Note that the pointwise $\lambda\to0$ limit of the integrands on the left-hand sides are just the ones on the right-hand sides. We want to apply \textit{Lebesgue's theorem} (see Appendix~\ref{ss:app:lebesgue} above) so that we can interchange the limit and the integral. To this end, we need a ``dominant'' function, i.e., an integrable function whose modulus is greater or equal to that of the integrand, independently of $\lambda$, for any $\lambda\,{>}\,0$. We do not explicitly write up this function, however, we give the various estimations that are needed to construct this function.
\begin{enumerate}
\item
The $f_s$ function is bounded everywhere, $f_s(q)\,{\le}\,K$. Also, the main point of introducing the subtraction of $f_s(0)$ and $f_s(2k)$ values as in \Eqs{e:app:finallimitD1s}{e:app:finallimitD2s} is that for the fractions appearing from these can be dominated by a function that is integrable around $q=0$ and $q=2k$ if $f_s$ is continuously differentiable everywhere: in this case for a finite $q_{\m{max}}$ value, 
\begin{align}
\label{e:fsfracbounded}
\frac{|f(q)\,{-}\,f(0)|}{q},\qquad\textnormal{and}\quad \frac{|f(q)\,{-}\,f(2k)|}{q\,{-}\,2k}\quad\textnormal{are bounded on $[0,q_{\m{max}}]$,}
\end{align}
and thus integrable here. Had we not have subtracted $f_s(0)$ and $f_s(2k)$, the fractions $\frac{f_s(q)}q$ and $\frac{f_s(q)}{q{-}2k}$ would have not be integrable around $q\,{=}\,0$ and $q\,{=}\,2k$. 

In the case of our preeminent practical example of source functions, the L\'evy distribution, $f_s(\bs q)$ is \textit{not} necessarily differentiable around $\bs q\,{=}\,0$. However, for our estimation to find the dominant function here, it is enough if the fractions written up in \Eq{e:fsfracbounded} are not bounded but can be dominated by an integrable function. This is readily satisfied also in the case of L\'evy distributions, where the following estimation is true instead:
\begin{align}
\label{e:powerpeak}
\textnormal{For L\'evy distributions, for any $\bs q$ and $\bs q'$,}\qquad |f(\bs q)\,{-}\,f(\bs q')|\le K'|\bs q\,{-}\,\bs q'|^\alpha;\quad\textnormal{this is enough.}
\end{align}
\item 
For the various factors appearing, the following estimations can be made. For real $X$ and $Y$, the pure imaginary power $X^{iY}$ is bounded by $e^{-\pi Y}$ from below and by $e^{\pi Y}$ from above. The $\c F_+(x)$ hypergeometric function is bounded on the whole complex $x$ plane.%
\footnote{
Recall that if $f(x)$ is an \textit{entire} (i.e., everywhere differentiable) non-constant function of the \textit{complex} $x$ variable, then by virtue of \textit{Liouville's theorem}, it cannot be a bounded function. Our $\c F_+(x)$ is not an entire function (owing to its branch cut on $x\,{\in}\,[1,\infty]$). The boundedness of $\c F_+(x)$ can be proven by considering various linear transformations of variables well known in the theory of hypergeometric functions; see Ref.~\cite{NIST:DLMF} for details.
}
\item 
Owing to the continuous differentiability of $\c F_+(x)$ around $x\,{=}\,0$, there exists a $C'$ constant for which $|\c F_+(x)\,{-}\,1|\,{\le}\,C'|x|$ for small enough $x$. Also, for pure imaginary powers, the following estimation can be made:
if $Y\in\mathbb R$, $0\,{\le}\,R\,{<}\,1$ and $|z|\,{<}\,R$, then $\big|(1{+}z)^{iY}\,{-}\,1\big|\,{\le}\,\rec{1-R}e^{\pi|Y|}\cdot|Y|\cdot|z|$ holds.%
\footnote{
The direction of this inequality is exactly the opposite of a similarly looking (but completely different) inequality, known as \textit{Bernoulli's inequality}, valid for the case of real exponents.
}
\end{enumerate}
From these, with some effort, one can construct the dominant function; this is best done by separately treating the $q\,{\in}\,[0,Q_{\m{max}}]$ and the $q\,{\in}\,[Q_{\m{max}},\infty]$ intervals, where $Q_{\m{max}}$ is an arbitrary value that is at a safe distance above $2k$; say: $Q_{\m{max}}\,{=}\,4k$. In this way, the exchange of the integral and the limit in \Eqs{e:app:finallimitD1s}{e:app:finallimitD2s} becomes justified, which leads to the desired results for $\c A_{1s}$ and $\c A_{2s}$. 

} 

\bibliographystyle{epj}
\bibliography{references}

\begin{thebibliography}{34}

\bibitem{STAR:2005gfr}
J.~Adams et~al. (STAR), Nucl. Phys. A \textbf{757}, 102 (2005), \texttt
  arXiv:{nucl-ex/0501009}

\bibitem{Adcox:2004mh}
K.~Adcox et~al. (PHENIX Coll.), Nucl.Phys. \textbf{A757}, 184 (2005), \texttt
  arXiv:{nucl-ex/0410003}

\bibitem{PHOBOS:2004zne}
B.B. Back et~al. (PHOBOS), Nucl. Phys. A \textbf{757}, 28 (2005), \texttt
  arXiv:{nucl-ex/0410022}

\bibitem{BRAHMS:2004adc}
I.~Arsene et~al. (BRAHMS), Nucl. Phys. A \textbf{757}, 1 (2005), \texttt
  arXiv:{nucl-ex/0410020}

\bibitem{HanburyBrown:1956bqd}
R.~Hanbury~Brown, R.Q. Twiss, Nature \textbf{178}, 1046 (1956)

\bibitem{Lisa:2005dd}
M.A. Lisa, S.~Pratt, R.~Soltz, U.~Wiedemann, Ann. Rev. Nucl. Part. Sci.
  \textbf{55}, 357 (2005), \texttt arXiv:{nucl-ex/0505014}

\bibitem{Goldhaber:1960sf}
G.~Goldhaber, S.~Goldhaber, W.Y. Lee, A.~Pais, Phys. Rev. \textbf{120}, 300
  (1960)

\bibitem{PHENIX:2004yan}
S.S. Adler et~al. (PHENIX), Phys. Rev. Lett. \textbf{93}, 152302 (2004),
  \texttt arXiv:{nucl-ex/0401003}

\bibitem{STAR:2004qya}
J.~Adams et~al. (STAR), Phys. Rev. C \textbf{71}, 044906 (2005), \texttt
  arXiv:{nucl-ex/0411036}

\bibitem{Lisa:2000ip}
M.A. Lisa, U.W. Heinz, U.A. Wiedemann, Phys. Lett. B \textbf{489}, 287 (2000),
  \texttt arXiv:{nucl-th/0003022}

\bibitem{Csanad:2008af}
M.~Csan\'ad, B.~Tomasik, T.~Cs\"{o}rg\H{o}, Eur. Phys. J. A \textbf{37}, 111 (2008),
  \texttt arXiv:{0801.4434}

\bibitem{Pratt:2008qv}
S.~Pratt, Phys. Rev. Lett. \textbf{102}, 232301 (2009), \texttt
  arXiv:{0811.3363}

\bibitem{PHENIX:2006nml}
S.S. Adler et~al. (PHENIX), Phys. Rev. Lett. \textbf{98}, 132301 (2007),
  \texttt arXiv:{nucl-ex/0605032}

\bibitem{Kisiel:2009iw}
A.~Kisiel, D.A. Brown, Phys. Rev. C \textbf{80}, 064911 (2009), \texttt
  arXiv:{0901.3527}

\bibitem{PHENIX:2017ino}
A.~Adare et~al. (PHENIX), Phys. Rev. C \textbf{97}, 064911 (2018), \texttt
  arXiv:{1709.05649}

\bibitem{NA61SHINE:2023qzr}
H.~Adhikary et~al. (NA61/SHINE) (2023), \texttt arXiv:{2302.04593}

\bibitem{CMS:2023xyd}
A.~Tumasyan et~al. (CMS) (2023), \texttt arXiv:{2306.11574}

\bibitem{Kincses:2019rug}
D.~Kincses, M.I. Nagy, M.~Csan\'ad, Phys. Rev. C \textbf{102}, 064912 (2020),
  \texttt arXiv:{1912.01381}

\bibitem{CoulCorrLevyIntegral}
\emph{{Correlation function calculation with L\'evy source and Coulomb FSI}}
  (2023), \texttt{https://https://github.com/csanadm/CoulCorrLevyIntegral}

\bibitem{Yano:1978gk}
F.B. Yano, S.E. Koonin, Phys. Lett. B \textbf{78}, 556 (1978)

\bibitem{Pratt:1997pw}
S.~Pratt, Phys. Rev. C \textbf{56}, 1095 (1997)

\bibitem{Rischke:1996em}
D.H. Rischke, M.~Gyulassy, Nucl. Phys. A \textbf{608}, 479 (1996), \texttt
  arXiv:{nucl-th/9606039}

\bibitem{Landau:1991wop}
L.D. Landau, E.M. Lifshits, \emph{{Quantum Mechanics}: {Non-Relativistic
  Theory}}, Vol. v.3 of \emph{Course of Theoretical Physics}
  (Butterworth-Heinemann, Oxford, 1991), ISBN 978-0-7506-3539-4

\bibitem{Csorgo:2003uv}
T.~Cs\"{o}rg\H{o}, S.~Hegyi, W.A. Zajc, Eur. Phys. J. C \textbf{36}, 67 (2004), \texttt
  arXiv:{nucl-th/0310042}

\bibitem{Csorgo:2004sr}
T.~Cs\"{o}rg\H{o}, S.~Hegyi, T.~Novak, W.A. Zajc, Acta Phys. Polon. B \textbf{36}, 329
  (2005), \texttt arXiv:{hep-ph/0412243}

\bibitem{Csorgo:2005it}
T.~Cs\"{o}rg\H{o}, S.~Hegyi, T.~Novak, W.A. Zajc, AIP Conf. Proc. \textbf{828}, 525
  (2006), \texttt arXiv:{nucl-th/0512060}

\bibitem{Cimerman:2019hva}
J.~Cimerma\v{n}, C.~Plumberg, B.~Tom\'a\v{s}ik, Phys. Part. Nucl. \textbf{51},
  282 (2020), \texttt arXiv:{1909.07998}

\bibitem{Csanad:2007fr}
M.~Csan\'ad, T.~Cs\"{o}rg\H{o}, M.~Nagy, Braz. J. Phys. \textbf{37}, 1002 (2007), \texttt
  arXiv:{hep-ph/0702032}

\bibitem{Kincses:2022eqq}
D.~Kincses, M.~Stefaniak, M.~Csan\'ad, Entropy \textbf{24}, 308 (2022), \texttt
  arXiv:{2201.07962}

\bibitem{Csanad:2019cns}
M.~Csan\'ad, S.~L\"ok\"os, M.~Nagy, Universe \textbf{5}, 133 (2019), \texttt
  arXiv:{1905.09714}

\bibitem{Nordsieck:1954zz}
A.~Nordsieck, Phys. Rev. \textbf{93}, 785 (1954)

\bibitem{NIST:DLMF}
{\relax DLMF}, \emph{{\it NIST Digital Library of Mathematical Functions}},
  \url{https://dlmf.nist.gov/}, Release 1.1.10 of 2023-06-15, f.~W.~J. Olver,
  A.~B. {Olde Daalhuis}, D.~W. Lozier, B.~I. Schneider, R.~F. Boisvert, C.~W.
  Clark, B.~R. Miller, B.~V. Saunders, H.~S. Cohl, and M.~A. McClain, eds.,
  \texttt{https://dlmf.nist.gov/}

\bibitem{davis2014methods}
P.~Davis, P.~Rabinowitz, W.~Rheinbolt, \emph{Methods of Numerical Integration},
  Computer Science and Applied Mathematics (Elsevier Science, 2014), ISBN
  9781483264288, \texttt{https://books.google.com/books?id=mbLiBQAAQBAJ}

\bibitem{rudin86real}
W.~Rudin, \emph{Real and Complex Analysis} (McGraw-Hill
  Science/Engineering/Math, 1986), ISBN 0070542341,
  \texttt{http://www.amazon.com/exec/obidos/redirect?tag=citeulike07-20\&path=ASIN/0070542341}

\end{thebibliography}

\end{document}